\documentclass[12pt]{article}

\usepackage{verbatim,color,amssymb}
\usepackage{amsmath}					
\usepackage{amsthm}					
\usepackage{natbib}
\usepackage{multirow}
\usepackage{setspace}
\usepackage[mathscr]{euscript}
\usepackage{fancyhdr}
\usepackage{enumitem}
\usepackage{graphicx}
\usepackage{lineno}
\usepackage{array,booktabs}
\usepackage{comment}

\usepackage{multibib}
\newcites{latex}{References}

\usepackage{lscape}

\usepackage{subfig}
\usepackage{setspace}
\usepackage{tikz}
\usetikzlibrary{arrows}
\usepackage{multirow}

\newtheorem{Thm}{\underline{\bf Theorem}}
\newtheorem{Assmp}{\underline{\bf Assumptions}}

\newtheorem*{Proof*}{Proof}

\newtheorem{Rem}{\underline{\bf Remark}}

\def\cC{\mathbb{C}}

\def\eE{\mathbb{E}}

\def\F{{\cal F}}

\def\H{{\cal H}}

\def\O{{\cal O}}

\def\U{{\cal U}}
\def\W{{\cal W}}

\def\X{{\cal X}}

\def\diag{\hbox{diag}}

\def\wh{\widehat}
\def\wt{\widetilde}

\def\diag{\hbox{diag}}
\def\log{\hbox{log}}

\def\var{\hbox{var}}
\def\cov{\hbox{cov}}

\def\trace{\hbox{trace}}

\def\etr{\hbox{etr}}

\def\Dir{\hbox{Dir}}

\def\Ga{\hbox{Ga}}

\def\MVN{\hbox{MVN}}

\def\Normal{\hbox{Normal}}
\def\MvMF{\hbox{MvMF}}
\def\TN{\hbox{TN}}
\def\Unif{\hbox{Unif}}
\def\vMF{\hbox{vMF}}

\def\P_25_ICML{{\it Proceedings of the 25th international conference on Machine learning}}

\def\bse{\begin{eqnarray*}}
\def\ese{\end{eqnarray*}}
\def\be{\begin{eqnarray}}
\def\ee{\end{eqnarray}}
\def\bq{\begin{equation}}
\def\eq{\end{equation}}

\def\wh{\widehat}

\def\trans{^{\rm T}}

\def\th{^{th}}
\def\bone{{\mathbf 1}}

\def\b1e{{\mathbf e}}
\def\ba{{\mathbf a}}

\def\bb{{\mathbf b}}

\def\bc{{\mathbf c}}
\def\bC{{\mathbf C}}

\def\bFf{{\mathbf f}}
\def\bF{{\mathbf F}}

\def\bH{{\mathbf H}}
\def\bh{{\mathbf h}}
\def\bI{{\mathbf I}}

\def\bM{{\mathbf M}}

\def\bP{{\mathbf P}}
\def\bq{{\mathbf q}}
\def\bQ{{\mathbf Q}}
\def\bR{{\mathbf R}}
\def\bs{{\mathbf s}}
\def\bS{{\mathbf S}}
\def\bt{{\mathbf t}}
\def\bT{{\mathbf T}}
\def\bu{{\mathbf u}}
\def\bU{{\mathbf U}}
\def\bv{{\mathbf v}}
\def\bV{{\mathbf V}}
\def\bw{{\mathbf w}}
\def\bW{{\mathbf W}}
\def\bx{{\mathbf x}}
\def\bX{{\mathbf X}}
\def\by{{\mathbf y}}
\def\bY{{\mathbf Y}}
\def\bz{{\mathbf z}}
\def\bZ{{\mathbf Z}}
\def\bS{{\mathbf S}}
\def\bzero{{\mathbf 0}}

\newcommand{\bmu}{\mbox{\boldmath $\mu$}}

\newcommand{\bpi}{\mbox{\boldmath $\pi$}}

\newcommand{\bvartheta}{\mbox{\boldmath $\vartheta$}}
\newcommand{\bepsilon}{\mbox{\boldmath $\epsilon$}}
\newcommand{\btheta}{\mbox{\boldmath $\theta$}}
\newcommand{\bbeta}{\mbox{\boldmath $\beta$}}

\newcommand{\bzeta}{\mbox{\boldmath $\zeta$}}
\newcommand{\bsigma}{\mbox{\boldmath $\sigma$}}
\newcommand{\bSigma}{\mbox{\boldmath $\Sigma$}}

\newcommand{\bLambda}{\mbox{\boldmath $\Lambda$}}

\newcommand{\norm}[1]{\left\Vert#1\right\Vert}

\renewcommand\footnoterule{\kern-3pt \hrule \textwidth 2in \kern 2.6pt}

\def\colred#1{\textcolor{red}{#1}}

\def\boxit#1{\vbox{\hrule\hbox{\vrule\kern6pt \vbox{\kern6pt \textcolor{blue}{#1}\kern6pt}\kern6pt\vrule}\hrule}}

\def\authorfootnote#1{{\let\thefootnote\relax\footnotetext{#1}}}

\allowdisplaybreaks

\begin{document}
\thispagestyle{empty}
\baselineskip=28pt

\begin{center}
{\LARGE{\bf 
Bayesian Semiparametric\\ 
Multivariate 
Density Deconvolution 
via Stochastic Rotation of Replicates
}}
\end{center}
\baselineskip=12pt

\vskip 2mm
\begin{center}
Arkaprava Roy\\
arkaprava.roy@ufl.edu\\
Department of Biostatistics,
University of Florida\\
2004 Mowry Road, Gainesville, FL  32611, USA\\
\vskip 2mm%
Abhra Sarkar\\
abhra.sarkar@utexas.edu \\
Department of Statistics and Data Sciences,
The University of Texas at Austin\\
2317 Speedway D9800, Austin, TX 78712-1823, USA\\
\end{center}

\vskip 8mm
\begin{center}
{\Large{\bf Abstract}} 
\end{center}

In multivariate density deconvolution, the distribution of a random vector needs to be estimated from replicates contaminated with measurement errors. 
This article presents a novel approach to multivariate deconvolution by stochastically rotating the replicates toward the corresponding true latent values. 
The method further accommodates conditionally heteroscedastic measurement errors  commonly observed in many real data applications. 
The estimation and inference schemes are developed within a Bayesian framework 
implemented via an efficient Markov chain Monte Carlo algorithm, 
appropriately accommodating uncertainty in all aspects of the analysis. 
The method's efficacy is demonstrated empirically through simulation experiments 
and practically in estimating the long-term joint average intakes of different dietary components from their measurement error-contaminated 24-hour dietary recalls. 

\baselineskip=12pt

\vskip 8mm
\baselineskip=12pt
\noindent\underline{\bf Some Key Words}: 
Copula,
Density deconvolution, 
Hamiltonian Monte Carlo,
Measurement error,
Markov chain Monte Carlo,
Nutritional epidemiology, 
Stiefel manifold,
Stochastic rotation,
von Mises-Fisher distribution.

\par\medskip\noindent
\underline{\bf Short/Running Title}: Multivariate Deconvolution via Stochastic Rotation

\par\medskip\noindent

\clearpage\pagebreak\newpage
\pagenumbering{arabic}
\newlength{\gnat}
\setlength{\gnat}{16pt}
\baselineskip=\gnat


\section{Introduction}
\label{sec: intro}
Many practical applications require the estimation of the unknown density of a
vector-valued random variable $\bx$. 
The variable $\bx$, however, may not be observed precisely, rather surrogate replicates $\bw$ contaminated with measurement errors $\bu$ may only be available. 
The replicates $\bw$ are then generated from a convolution of the density of $\bx$ and the density of
the measurement errors $\bu$, 
and the problem of estimating the density of $\bx$ from available
contaminated measurements $\bw$ becomes a problem of multivariate deconvolution of densities.
Such problems routinely arise in diverse application areas, including especially in nutritional epidemiology \citep{bennett2017systematic,keogh2020stratos1,shaw2020stratos2}, 
where the problem of estimating long-term average intakes of different dietary components from their error-contaminated 24-hour recalls is of fundamental importance.

This article proposes a robust approach to multivariate density deconvolution 
in the presence of conditionally heteroscedastic errors $\bu$ from an unknown probability law, 
relying on the idea of probabilistically rotating the replicates $\bw$ toward the underlying latent $\bx$ in a statistically principled manner.

Throughout this article, 
for random vectors $\bs$ and $\bt$, we denote the marginal density of $\bs$, 
the joint density of $(\bs,\bt)$, 
and the conditional density of $\bs$ given $\bt$, 
by the generic notation $f_{\bs}, f_{\bs,\bt}$ and $f_{\bs\vert \bt}$, respectively.
Likewise, for univariate random variables $s$ and $t$, the corresponding densities are denoted by $f_{s},f_{s,t}$ and $f_{s\vert t}$, respectively.

The literature on density deconvolution is really vast \citep{Carroll2006,Buonaccorsi2010}. 
The early literature focused primarily on univariate problems with a single contaminated measurement for each subject and the measurement errors 
independently and identically distributed according to some known probability law $f_{u}$, often normal.
Deconvoluting kernel-based approaches have been studied by
\cite{stefanski1990deconvolving,devroye1989consistent,fan1991global,fan1991optimal}, among others. 
See also \cite{madrid2018deconvolution,newton2002nonparametric}.  
The distribution of measurement errors is, however, rarely known in practice. 
Robust deconvolution methods 
with the unknown aspects of the error density estimated using replicated proxies $w$ for the unknown values of $x$  
have thus been considered \citep[and others]{LiVuong1998,digglehall:1993,Delaigle2008}.
A Bayesian likelihood-based approach with an unknown but symmetric unimodal density $f_{x}$ has recently been developed in \cite{su2020nonparametric}.

The assumption of independence of $u$ from $x$ is also often highly impractical, especially in nutritional epidemiology applications, 
where patterns of conditional heteroscedasticity can be very prominently seen. 
Bayesian hierarchical frameworks and associated Markov chain Monte Carlo (MCMC) based computational machinery 
have recently been shown to provide powerful tools for solving complex deconvolution problems under more realistic scenarios, 
including when the measurement error distribution can be asymmetric, heavy-tailed, conditionally heteroscedastic, etc. 
\citep{Staudenmayer_etal:2008,sarkar2014bayesian,sarkar2018bayesian,sarkar2021bayesian}.  
In their seminal work, \cite{Staudenmayer_etal:2008} assumed the  errors $u$ to be normally distributed but allowed the variability of $u$ to depend on $x$, 
employing positive mixtures of B-splines to flexibly characterize both $f_{x}$ and the conditional variability $\var(u \vert x)$.
\cite{sarkar2014bayesian} further relaxed the assumption of normality of $u$, 
employing flexible mixtures of normals \citep{Escobar_West:1995, fruhwirth2006finite} to model both $f_{x}$ and $f_{u\vert x}$.

In stark contrast to the univariate setting, the multivariate problem has garnered little attention in the literature. 
\cite{masry1991multivariate,youndje2008optimal,comte2013anisotropic,hazelton2009nonparametric,hazelton2010semiparametric,eckle2017multiscale}
considered scenarios with errors $\bu$  
from a known probability law, independent from $\bx$.
\cite{bovy2011extreme} obtained a Bayesian maximum-\emph{a-posteriori} estimate of $f_{\bx}$ modeled by flexible mixtures of multivariate normal kernels, assuming $f_{\bu}$ to be multivariate normal, independent from $\bx$ with subject-specific but known covariance matrices.
Utilizing the flexibility of Bayesian hierarchical frameworks,  \cite{sarkar2018bayesian,sarkar2021bayesian} developed robust multivariate deconvolution methods, 
relaxing the restrictive assumptions of known error probability laws, homoscedasticity, and independence from $\bx$, etc. 
\cite{sarkar2018bayesian} modeled $f_{\bx}$ and $f_{\bu\vert \bx}$ using flexible mixtures of multivariate normals 
whereas \cite{sarkar2021bayesian} adopted a complementary approach, modeling the marginals $f_{x_{\ell}}$ and $f_{u_{\ell}\vert x_{\ell}}$ first 
and then building the joint distributions $f_{\bx}$ and $f_{\bu\vert\bx}$ by modeling the dependence structures using Gaussian copulas.

The focus of this article is also on multivariate deconvolution with conditionally heteroscedastic measurement errors from an unknown distribution in the presence of replicated proxies for each subject. 
To that end, we propose a novel approach to multivariate density deconvolution that assumes 
the replicates $\bw$ to be generated by first stochastically rotating the underlying true $\bx$ and then stochastically stretching or contracting their lengths. 
This is achieved by multiplying each $\bx$ first with an orthogonal rotation matrix $\bQ$ 
and then with a scalar length adjustment factor $r$. 
Going a significant step further, we also accommodate conditional heteroscedasticity by allowing the distributions of both the rotation matrices and the length-adjusting factors 
to flexibly depend on the latent true $\bx$'s. 
The conditional distributions $f_{\bu \vert \bx}$, and hence $f_{\bw \vert \bx}$, are then obtained as novel functions of the $\bQ$'s. 
For the main density of interest $f_{\bx}$, we adopt a copula-based approach with the marginals modeled by flexible mixtures of truncated normals with shared atoms as in \cite{sarkar2021bayesian}.
We take a Bayesian route to estimation and inference, implemented via an efficient MCMC algorithm, 
appropriately accommodating uncertainty in all aspects of our analysis. 
We illustrate our method's empirical efficacy through simulation experiments. 
Its practical utility is demonstrated in nutritional epidemiology applications in estimating the long-term joint average intakes of different dietary components from their measurement error-contaminated 24-hour dietary recalls.

Traditionally, the literature on deconvolution almost exclusively assumes the measurement errors to be additive. 
In Section \ref{sec: classical} of this paper, we show that our rotation-based model can be reformulated as a classical additive model. 
What the rotation-based view does still is to provide a new perspective on measurement errors leading to a new way of constructing the likelihood function and resulting in new algorithms for inference.

Rotation-guided modeling of multivariate data is indeed getting increasing popularity in statistics \citep{hoff2009hierarchical,hoff2009simulation, mccormick2015latent, Minerva2020, song2022curved}. 
In the deconvolution literature, stochastic rotations have been proposed for directional data in \cite{kim1998deconvolution}, where the latent objects of primary interest, as well as the observed replicates, were all orthogonal matrices, with additional theoretical insights into this setup provided in \cite{kim2001deconvolution}. 
Our work, however, is focused on Euclidean deconvolution problems in the presence of replicates contaminated with conditionally heteroscedastic measurement errors. 
The deconvolution problem we consider and the solution we propose are thus very different from \cite{kim1998deconvolution}.


{Overall, this article makes several important contributions to the literature on multivariate density deconvolution - 
(a) we introduce a new framework for multivariate deconvolution via stochastic rotation of the error-contaminated replicates toward their underlying true values, 
(b) additionally, we also address the significantly challenging problem of accommodating conditionally heteroscedastic errors in this newly introduced framework, and 
(c) we introduce HMC-based advanced MCMC methods to the deconvolution problem, significantly improving computational efficiency.}

The rest of this article is organized as follows. 
Section \ref{sec: prelims} presents some important preliminary results used in the construction of our likelihood function. 
Section \ref{sec: models} details our proposed stochastic rotation-based approach to multivariate deconvolution, 
including 
likelihood construction, 
prior specification, and 
outline of posterior computation. 
Section \ref{sec: sim studies} presents the results of some simulation experiments, illustrating the method's empirical performances. 
Section \ref{sec: applications} presents the results produced by the proposed method applied to the problem of estimating the true long-term average intakes of different dietary components from their measurement error-contaminated 24-hour recalls. 
Section \ref{sec: discussion} contains concluding remarks. 
Substantive additional details are presented in the supplementary materials.

\section{Preliminaries} \label{sec: prelims}
\subsection{Geometry of Vector Rotations}
To motivate our modeling framework, we first discuss some geometric properties of vector rotations. 
For any two vectors $\ba\in\mathbb{R}^d$ and $\bb\in\mathbb{R}^d$, there exists a $d\times d$ orthonormal rotation matrix $\bQ_{ab}$ such that $\frac{\ba}{\|\ba\|_{2}}=\bQ_{ab}\frac{\bb}{\|\bb\|_{2}}$, where $\|\cdot\|_{2}$ stands for the Euclidean norm. 
We can thus establish that 
\be
 \ba=s\bQ_{ab}\bb,\label{eq:ortho}  
\ee
for the scalar $s={\|\ba\|_{2}}/{\|\bb\|_{2}}$. Geometrically speaking, the orthonormal matrix $\bQ_{ab}$ rotates the unit vector $\frac{\bb}{\|\bb\|_{2}}$ towards the unit vector $\frac{\ba}{\|\ba\|_{2}}$ and the scalar $s$ takes care of the change in magnitude due to this transformation from $\bb$ to $\ba$.
Let $\wt{\ba}=\ba/\|\ba\|_{2}$ and $\wt{\bb}=\bb/\|\bb\|_{2}$.
Then, using the Householder reflection result \citep{householder1958unitary} for total internal reflection under Snell's law, 
a solution for $\bQ_{ab}$ is $\bI_{d}-2\bv\bv\trans/\|\bv\|_{2}^{2}$, where $\bv=\wt{\ba}-\wt{\bb}$.
However, since it is based on a single Householder transformation, the solution is always symmetric, 
hence making it difficult to impose any distributional assumption.
We thus consider a different solution based on a frequently used technique in numerical analysis to compute the orthogonal component $Q$ in the $QR$-decomposition of a matrix \citep{stewart1980efficient}.
The exact expression of $\bQ_{ab}$ is established in the following theorem. 

\begin{Thm}
\label{thm:orthosol}
For any $i=1,\dots,d$, let $\ba_{i}=\ba+\|\ba\|_{2} \b1e_{i}$ and $\bb_{i}=\bb+\|\bb\|_{2} \b1e_{i}$, where $\b1e_{i}$ is the unit vector with $1$ at the $i\th$ place. 
Then, for $\mathcal{H}(\ba)=\bI_{d}-2\ba_{i}\ba_{i}\trans/\|\ba_{i}\|_{2}^{2}$ and $\mathcal{H}(\bb)=\bI_{d}-2\bb_{i}\bb_{i}\trans/\|\bb_{i}\|_{2}^{2}$, $\bQ_{ab}=\mathcal{G}(\ba, \bb)=\H(\ba)\H(\bb)$ satisfies (\ref{eq:ortho}).
\end{Thm}

\begin{figure}
\centering
\vskip -5pt
\includegraphics[height=7cm, trim=1cm 1.2cm 1cm 1cm, clip=true]{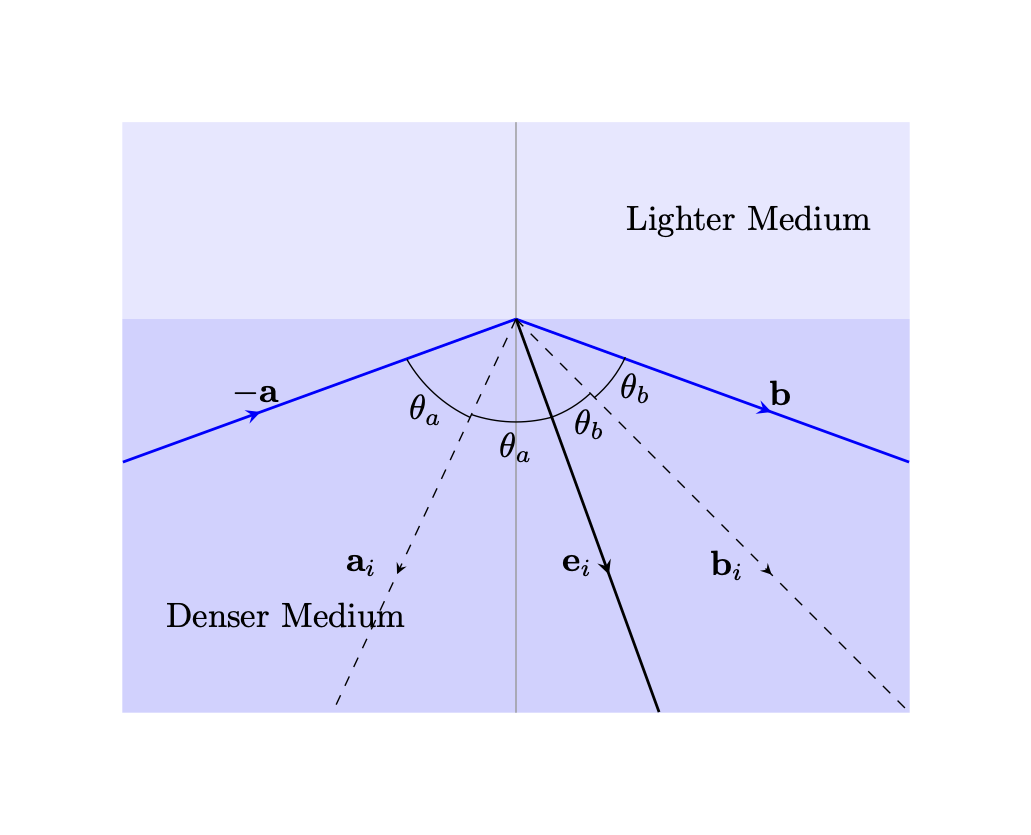}
\vskip -0pt
\caption{Rotation by total internal reflection via Householder transformations.}
\label{fig: reflection}
\vskip -5pt
\end{figure}

The result above is crucial in computing the likelihood function of our proposed model. 
The proof is based on some results from Euclidean geometry. 
If the singular value decomposition of $\ba\bb\trans$ is $\bU\bSigma\bV\trans$ with the singular values in $\bSigma=\diag(\sigma_{1,1},\dots,\sigma_{d,d})$, then a solution for $\bQ_{ab}$ is the orthogonal matrix $\bU\bV\trans$. 
The matrix $\ba\bb\trans$ has only one non-zero singular value, $\|\ba\|_{2}\|\bb\|_{2}$, which we can assume to be $\sigma_{i,i}$ without any loss of generality.
Based on a similar reflection argument as before, we can then compute $\bU$ and $\bV$ as functions of $\ba$ and $\bb$ explicitly. 
Specifically, we take $\ba_{i}=\ba+\|\ba\|_{2} \b1e_{i}$, the bisector of the angle between $\ba$ and the $i\th$ unit vector $\b1e_{i}$ on the reflecting surface. 
Then the Householder transformation matrix based on this bisector, $\bI_{d}-2\ba_{i}\ba_{i}\trans/\|\ba_{i}\|_{2}^{2}$, is a possible solution for $\bU$ as this Householder reflection will transform $\ba$ to $-\|\ba\|_{2} \b1e_{i}$ and thus $\bU\ba=-\|\ba\|_{2}\b1e_{i}$. 
It can be verified easily by noting that $\ba_{i}\trans\ba=\|\ba\|_{2}\ba_{i}\trans\b1e_{i}$.
Similarly, we can compute $\bV$ from $\bb$. 
With $\mathcal{H}(\ba)=\bI_{d}-2\ba_{i}\ba_{i}\trans/\|\ba_{i}\|_{2}^{2}$ and $\mathcal{H}(\bb)=\bI_{d}-2\bb_{i}\bb_{i}\trans/\|\bb_{i}\|_{2}^{2}$, 
we then have $\bQ_{ab}=\mathcal{G}(\ba, \bb)=\H(\ba)\H(\bb)$ (Figure \ref{fig: reflection}).

\begin{Rem}
It is easy to check that $\H(\ba)\H(\bb)=\bI_d$ if and only if $\ba/\|\ba\|_{2}=\bb/\|\bb\|_{2}$. 
\end{Rem}

For the rest of the paper, 
without loss of generality, 
we use the result of Theorem \ref{thm:orthosol} with the first unit vector $\b1e_{i}=\b1e_{1}$. 
Numerical experiments with other choices produced near-identical results. 
In the next section, we use the result to compute the likelihood function of the replicates in our multivariate density deconvolution model introduced in the next section.


\subsection{von-Mises Fisher and  Matrix von-Mises Fisher Dstributions}

The von-Mises Fisher distribution (vMF) and the matrix von-Mises Fisher distribution (MvMF) play important roles in our model construction and computation. 
We thus provide a brief description of these distributions here, starting with the vMF first, for easy reference. 

The vMF distribution for a $d$-dimensional vector $\bw$ is defined as 
\begin{align}
    f(\bw) = C_{d}(c)\exp(c\bmu^T\bw), \quad \bw\in \{\bz:\|\bz\|_{2}=1\},
\end{align}
where $\bmu$ is a unit vector signifying the mean direction of $\bw$ 
and $c\geq0$ stands for a scalar concentration parameter. 
The normalizing constant $C_{d}(c)=\frac{c^{d/2-1}}{(2\pi)^{d/2}I_{d/2-1}(c)},$ where $I_{\nu}$ denotes the modified Bessel function of the first kind at order $\nu$. 
The maximum likelihood estimates for $\bmu$ and $c$ are obtained in \cite{jupp1979maximum}.

We next review the MvMF distribution for orthogonal matrices on the Stiefel manifold. 
The distribution was first proposed in \cite{downs1972orientation} and thoroughly studied in \cite{khatri1977mises}.
Since then, it has become a popular distributional choice for random matrices supported in the space of orthogonal matrices.

The MvMF distribution for a $q \times d$ matrix $\bQ = ((Q_{\ell_{1},\ell_{2}}))$, with $q \leq d$ and $\bQ\bQ\trans = \bI_{q}$, is defined as
\begin{align}
    f(\bQ) = \frac{1}{M(\bF)}\etr(\bF\bQ\trans), \quad \bQ\in \O_{q,d},
\end{align}
where $\etr(\cdot) = \exp\{\trace(\cdot)\}$, 
$\bF$ is $q\times d$ dimensional parameter matrix,
$\O_{q,d}$ is the set of all $q\times d$ orthogonal matrices 
(i.e., the $q\times d$ Stiefel manifold), 
and $M(\bF)$ is the normalization constant.
To keep the description simple, we only consider the case directly relevant to our deconvolution problem, namely $q=d$ with a diagonal $\bF$, and review some key results from \citelatex{khatri1977mises}.
The MGF of $\bQ$ is given by $\eE\{\etr(\bT\bQ\trans)\}=\frac{M(\bF+\bT)}{M(\bF)}$.
Let $\bT_{1}=\diag(t_{1,1},\ldots,t_{1,d})$ and $\bT_{2}=\diag(t_{2,1},\ldots,t_{2,d})$ be matrices such that $t_{1,ell}=1/t_{2,\ell}$ for all $\ell=1,\dots,d$.
Thus, the distributions of $\bZ=\bT_{1}\bQ \bT_{2}$ and $\bQ$ are identical when $\bF$ is diagonal. 
Hence, 
\bse
& \eE(Q_{\ell_{1},\ell_{2}})=\eE(z_{\ell_{1},\ell_{2}})\implies (1-t_{1,\ell_{1}}t_{2,\ell})\eE(Q_{\ell_{1},\ell_{2}})=0~~~\text{and}\\
& \eE(Q_{\ell_{1},\ell_{2}}Q_{\ell_{1}',\ell_{2}'})=\eE(z_{\ell_{1},\ell_{2}}z_{\ell_{1}',\ell_{2}'})\implies (1-t_{1,\ell_{1}}t_{2,\ell_{2}}t_{1,\ell_{1}'}t_{2,\ell_{2}'})\eE(Q_{\ell_{1},\ell_{2}}Q_{\ell_{1}',\ell_{2}'})=0.
\ese
This implies $\eE(Q_{\ell_{1},\ell_{2}})=0$ for all $k \neq \ell$ and $\eE(Q_{\ell_{1},\ell_{2}}Q_{\ell_{1}',\ell_{2}'})=0$ for either $\ell_{1} \neq \ell_{2}$ or $\ell_{1}' \neq \ell_{2}'$. 
To compute the expectation of the diagonal entries, we can take the derivative of the MGF with respect to $\bT$. 
Hence, we get $\eE(Q_{\ell_{1},\ell_{1}})=\frac{\partial \log M(\bF)}{\partial F_{\ell_{1},\ell_{1}}}$ and $\eE(Q_{\ell_{1},\ell_{1}}Q_{\ell_{2},\ell_{2}})=\frac{\partial \log M(\bF)}{\partial F_{\ell_{1},\ell_{1}}\partial F_{\ell_{2},\ell_{2}}}$.

\section{Deconvolution via Stochastic Rotation} 
\label{sec: models}
Our main objective is to estimate the density of a $d$-dimensional vector $\bx$. 
However, we do not have accurate measurements of $\bx$. 
For each unobserved $\bx_{i}$, we instead have $m_{i}$ replicated proxies $\bw_{i,j}$'s contaminated with some error where $i=1,\ldots,n$ and $j=1\ldots,m_{i}$, and $m_{i}\geq 3$ for all $i$. 
Each observation $\bw_{i,j}$ may be viewed as a point in the Cartesian co-ordinate system. 
A representation of $\bw_{i,j}$ can then  be obtained in terms of its norm $\|\bw_{i,j}\|_{2}$ and its direction from the origin  $\bw_{i,j}/\|\bw_{i,j}\|_{2}$.
In this paper, our characterization of $\bw_{i,j}$ takes inspiration from the above representation and the relation in~\eqref{eq:ortho}. 
Specifically, our model for the replicates $\bw_{i,j}$ conditional on the underlying true $\bx_{i}$ is
\begin{align}
\begin{split}
    &\bw_{i,j}=r_{i,j}\bC_{i}\bQ_{i,j}\bx_{i},\\
    & \bQ_{i,j}\sim\textrm{MvMF}(\bF_{i})\propto\etr(\bF_{i}\bQ_{i,j}\trans),\quad \bF_{i}=\textrm{diag}\{\kappa_{1}(x_{1,i}),\ldots, \kappa_{d}(x_{d,i})\},\\
    &\log(r_{i,j})\sim \Normal\{-s^{2}(\|\bx_{i}\|_{2}/d)/2,s^{2}(\|\bx_{i}\|_{2}/d)\},\\
    &\textstyle \kappa_{\ell}(x)=\sum_{k=1}^{K_{\kappa}}\beta_{\kappa,\ell,k}B_{k}(x),\quad s^{2}(\|\bx_{i}\|_{2}/d)=\sum_{k=1}^{K_{s}}\beta_{s,k}B_{k}(\|\bx_{i}\|_{2}/d).
\end{split}\label{eq:Orthmoddel}
\end{align}
Here $r_{i,j}$'s are scalars with $\eE(r_{i,j} \vert \bx_{i})=1$ and $\var(r_{i,j} \vert \bx_{i})=\exp\{s^{2}(\|\bx_{i}\|_{2}/d)\}-1$, 
$\bQ_{i,j} = ((Q_{\ell,\ell',i,j}))$'s are $d \times d$ rotation matrices, 
and $B_{k}(x)$ are B-spline bases spanning the interval $[A,B]$ \citep{de1978practical}. 
The $d \times d$ dimensional scaling matrices $\bC_{i}=((C_{\ell,\ell',i}))$ ensures that the replicates $\bw_{i,j}$'s are stochastically centered around the corresponding true $\bx_{i}$. 
Specifically, we set $\bC_{i} = \{\eE(\bQ_{i,j}\vert\bx_{i})\}^{-1}$, 
so that $\eE(\bw_{i,j}\vert\bx_{i}) = \eE(r_{i,j}|\bx_{i})\eE(\bC_{i}\bQ_{i,j}\vert\bx_{i})\bx_{i}=1\times\bI_{d}\times\bx_{i}=\bx_{i}$. 
As seen in Section \ref{sec: prelims}, for a diagonal $\bF_{i}$, the Euclidean expectation of $\bQ_{i,j}$ is also diagonal. 
Based on the results of \cite{khatri1977mises}, 
we specifically have $\eE(Q_{\ell,\ell',i,j} \vert \bx_{i}) =  C_{\ell,\ell,i}^{-1}=\frac{d\log\{M(\bF_{i})\}}{d\kappa_l}$. 
The matrices $\bF_{i}$ also determine how far the corresponding $\bQ_{i,j}$'s are allowed to vary around $\bI_{d}$, larger values of $\kappa_{\ell}(\cdot)$ inducing greater concentration of the $\bQ_{i,j}$'s around $\bI_{d}$.
Throughout the paper, we often keep the $\bx_{i}$'s implicit in $\bF_{i} = \bF_{i}(\bx_{i})$ and $\bC_{i}=\bC_{i}(\bF_{i}) = \bC_{i}(\bx_{i})$ to keep the notation simple.

\begin{figure}
\centering
\vskip -5pt
\includegraphics[height=7cm, trim=1cm 1.2cm 1cm 1cm, clip=true]{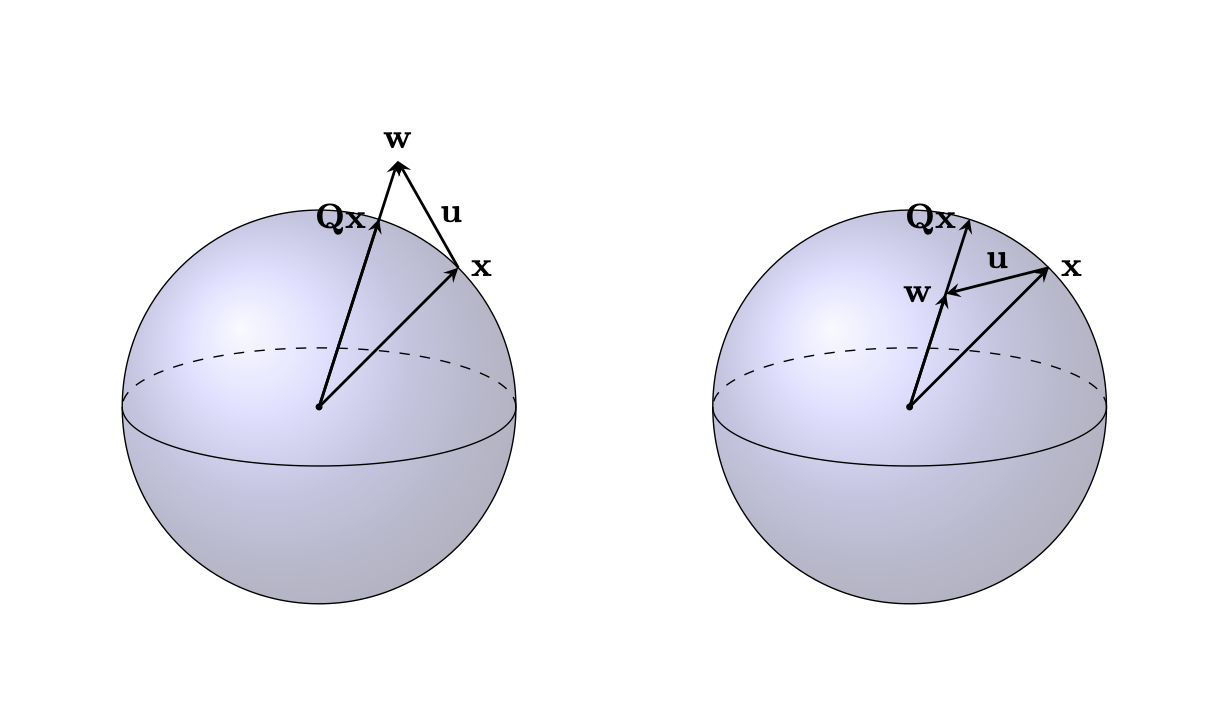}
\vskip -0pt
\caption{In the classical view of measurement error models, an observation $\bw$ for an underlying latent $\bx$ is obtained by contaminating $\bx$ with additive measurement error $\bu$. 
In the alternative view adopted in this article, $\bw$ is generated by first rotating $\bx$ to $\bQ\bx$ and then stretching (left panel) or contracting (right panel) it to $\bw$. 
The $3$-dimensional case is shown here, the shaded spheres representing the set of all vectors that can be generated by orthogonal rotations of $\bx$.}
\label{fig: Stochastic Rotation}
\vskip -5pt
\end{figure}

Rewriting the model as $\bC_{i}^{-1}\bw_{i,j}=r_{i,j}\bQ_{i.j}\bx_{i}$ and 
following Theorem \ref{thm:orthosol} in Section~\ref{sec: prelims}, 
the solutions for $\bQ_{i,j}$ and $r_{i,j}$ are $\mathcal{G}\left(\bC_{i}^{-1}\bw_{i,j},\bx_{i}\right)$ and $\|\bC_{i}^{-1}\bw_{i,j}\|_{2} / \|\bx_{i}\|_{2}$, respectively. 
The conditional likelihood of the replicates is therefore given by 
\bse
&\textstyle f_{\bw \vert \bx}(\bw_{i,j} \vert \bx_{i}) = 
\frac{\etr\left\{{\mathcal{G}\left(\bC_{i}^{-1}\bw_{i,j},\bx_{i}\right) \bF_{i}}\right\}}{M(\bF_{i})}       
\frac{\|\bx_{i}\|_{2}}{\sqrt{2\pi}\|\bC_{i}^{-1}\bw_{i,j}\|_{2} s(\|\bx_{i}\|_{2}/d)}\exp\left[-\frac{\left\{\log(\|\bC_{i}^{-1}\bw_{i,j}\|_{2}/\|\bx_{i}\|_{2}) + s^{2}(\|\bx_{i}\|_{2}/d)/2\right\}^{2}}{2s^{2}(\|\bx_{i}\|_{2}/d)}\right].
\ese

Although $M(\bF_{i})$ is computationally intractable, 
depending on the magnitudes of the elements in $\bF_{i}$, efficient approximations are available in \cite{khatri1977mises}. 
In most practical applications, including ours, the replicates $\bw_{i,j}$ are expected to lie in the general direction of the underlying $\bx_{i}$, 
which implies that the $\bQ_{i,j}$'s shouldn't deviate too far from the identity matrix $\bI_{d}$. 
Throughout this article, the functions $\kappa_{i}(\cdot)$'s are thus assumed to take large values which result in a higher concentration of $\bQ_{i,j}$ around $\bI_{d}$. 
Thus, it is reasonable to consider the following approximation from \cite{khatri1977mises} for large diagonal entries in $\bF_{i}$ 
\be
    M(\bF_{i})\approx\left\{\frac{2^{-\frac{1}{4}d(d+5)+\frac{1}{2}d^{2}}}{\pi^{\frac{d}{2}}}\right\}\etr(\bF_{i})\prod_{j=1}^{d}\Gamma\left(\frac{d-j+1}{2}\right)\left[\prod_{\ell'=2}^d\prod_{\ell<\ell'}\{\kappa_{\ell}(x_{\ell,i})+\kappa_{\ell'}(x_{\ell',i})\}^{\frac{1}{2}}\right]^{-1}. \label{eq:normconst}
\ee
In what follows, for brevity, sometimes we also use the notation $\bF_{i} = \diag(\bFf_{i})$ where $\bFf_{i}=(f_{1,i},\dots,f_{d,i})\trans= \{\kappa_{1}(x_{1,i}),\ldots,\kappa_d(x_{d,i})\}\trans$. 
Also, when we say $\bF_{i}$ is large, we mean its diagonal entries $\bFf_{i}$ are all large.

Following \cite{sarkar2021bayesian}, 
we model the joint density $f_{\bx}$ of $\bx$ using 
a Gaussian copula with component-wise univariate marginals characterized by flexible mixtures of truncated normals, 
truncated to their common support $[A,B]$. 
Specifically, we let 
\bse
    &f_{\bx}(\bx)=\bR_{\bx}^{-1}\exp\left\{-\by_{\bx}\trans(\bR_{\bx}^{-1}-\bI_{d})\by_{\bx}\right\}\prod_{\ell=1}^{d} f_{\bx,\ell}(x_{\ell}),\\
    & f_{\bx,\ell}(x_{\ell})=\sum_{k=1}^{K}\pi_{\ell,k}\TN(x_{\ell}\vert\mu_{k},\sigma_{k}^{2}, [A, B]),
\ese
where $\by_{\bx}=(y_{x,1},\dots,y_{x,d})\trans$ 
with $F_{x,\ell}(x_{\ell})=\Phi(y_{x,\ell})$ for each $\ell=1,\dots,d$, 
where $F_{x,{\ell}}$ is the cumulative distribution function (cdf) corresponding to $f_{x,{\ell}}$. 
Our model for the correlation matrix $\bR_{\bx}$ considers the spherical coordinate representation of Cholesky factorization. 
In this representation, the correlation matrix $\bR_{\bx}$ is written as $\bR_{\bx}=\bV\bV\trans$, 
where the $m$-$th$ row of $\bV$ is $\bV_{m,1}=\prod_{s=1}^{m-1}\sin(\zeta_{m,s})$, $\bV_{m,k}=\prod_{s=1}^{m-k-1}\sin(\zeta_{m,s})\cos(\zeta_{m,m-k-1})$ for $k=2,\ldots,m$. 
The rest of the entries are all zeros. 
Thus, $\bV$ is a lower triangular matrix. 
The angles $\zeta_{m,s}$ are supported on $[0,\pi]$ for $s< m-1$ and the $\zeta_{m,m-1}$'s are supported on $[0,2\pi]$.

{The truncated support $[A,B]$ for the marginals is consistent with modeling conditional heteroscedasticity  later in Section \ref{sec: cond heteroscedasiticity} by mixtures of B-splines which by definition have bounded local supports spanning a finite interval. 
Having a common unit free support $[A,B]$ for all components $\ell$ also greatly simplifies assigning priors on the parameters of the mixture components as well as the choices of these prior hyper-parameters. 
This is often easy to achieve in practice via simple linear transformations of the originally observed proxies. 
See, e.g., Section S.5. in the supplementary material.}

\subsection{Conditional Heteroscedasticity Characterization} \label{sec: cond heteroscedasiticity}
There are two different ways we accommodate conditional heteroscedasticity in the measurement errors in our model - (a) by allowing the distribution of length adjustment factors $r_{i,j}$ to depend on the underlying latent $\bx_{i}$ via the function $s^{2}(\|\bx_{i}\|_{2}/d)$, and (b) by allowing the $\kappa_{\ell}$ parameters to vary flexibly as functions of the corresponding latent $x_{\ell,i}$.
The part $s^{2}(\|\bx\|_{2}/d)$ accommodates our expectation that larger adjustments are needed for larger true latent vectors $\bx$;
whereas the functions $\kappa_{\ell}(x_{\ell,i})$'s accommodate the expectation that larger deviations of $w_{\ell,i,j}$ should be allowed around larger values of $x_{\ell,i}$. 
More specifically, we have 
$\cov(\bw_{i,j}\vert\bx_{i})=\cov(r_{i,j}\bC_{i}\bQ_{i,j}\bx_{i}\vert\bx_{i})=\eE\{\cov(r_{i,j}\bC_{i}\bQ_{i,j}\bx_{i}\vert\bx_{i}, r_{i,j})\}+\cov\{\eE(r_{i,j}\bC_{i}\bQ_{i,j}\bx_{i}\vert\bx_{i}, r_{i,j})\}
=\exp\{s^{2}(\|\bx_{i}\|_{2}/d)\}\var(\bC_{i}\bQ_{i,j}\bx_{i}\vert\bx_{i}) + [\exp\{s^{2}(\|\bx_{i}\|_{2}/d)\}-1]\bx_{i}\bx_{i}\trans$.
As seen in Section \ref{sec: prelims}, 
for a diagonal $\bF_{i}$, $\eE(Q_{\ell_{1},\ell_{2},i,j}Q_{\ell_{1}',\ell_{2}',i,j})=0$ except for $\ell_{1}=\ell_{2}=\ell_{1}'=\ell_{2}'$. 
Thus we have $\cov(\bC_{i}\bQ_{i,j}\bx_{i}\vert\bx_{i}) = \bC_{i}\diag(\bx_{i})\bV_{i}\diag(\bx_{i})\bC_{i}$, 
where $\bV_{i} = ((V_{\ell,\ell',i}))$ is the covariance matrix for the diagonal entries in $\bQ_{i,j}$. 
From equation (2.11) of \cite{khatri1977mises}, we have 
\bse
V_{\ell,\ell',i}=\frac{\partial^{2}}{\partial_{\ell}\partial_{\ell'}}\log[M\{\diag(\bFf_{i})\}].
\ese
For a large $\bF_{i}$, relying on the approximation of $M(\bF_{i})$, we can compute the entries $V_{\ell,\ell',i}$ as 
\bse
    V_{\ell,\ell',i} &\approx& \frac{1}{2(f_{\ell,i}+f_{\ell',i})^{2}} ~~~~~~~~\textrm{for}~ \ell\neq \ell',\\
    &\approx&\sum_{\ell'\neq \ell}\frac{1}{2(f_{\ell,i}+f_{\ell',i})^{2}} ~~~\textrm{for}~  \ell=\ell'.
\ese
To simplify notation, let us denote $\cov(\bw_{i,j}\vert\bx_{i})$ also by $\bS_{i}=((S_{\ell,\ell',i}))$. 
The other approximations are then 
\bse
    & C_{\ell,\ell,i} \approx 1\bigg/\left\{1-\sum_{\ell\neq \ell'}\frac{1}{2(f_{\ell,i}+f_{\ell',i})}\right\}, \\
    & S_{\ell,\ell',i} \approx \frac{\exp\{s^{2}(\|\bx_{i}\|_{2}/d)\}x_{\ell,i}x_{\ell',i}}{2(f_{\ell,i}+f_{\ell',i})^{2}}\bigg/\left[\left\{1-\sum_{\ell\neq \ell'}\frac{1}{2(f_{\ell,i}+f_{\ell',i})}\right\}\left\{1-\sum_{\ell\neq 
    \ell'}\frac{1}{2(f_{\ell,i}+f_{\ell',i})}\right\}\right] \\&\quad+ [\exp\{s^{2}(\|\bx_{i}\|_{2}/d)\}-1]x_{\ell,i}x_{\ell',i}, \\
    & S_{\ell,\ell,i} \approx \sum_{\ell\neq \ell'}\frac{\exp\{s^{2}(\|\bx_{i}\|_{2}/d)\}x_{\ell,i}^{2}}{2(f_{\ell,i}+f_{\ell',i})^{2}}\bigg/\left\{1-\sum_{\ell\neq \ell'}\frac{1}{2(f_{\ell,i}+f_{\ell',i})}\right\}^{2} + [\exp\{s^{2}(\|\bx_{i}\|_{2}/d)\}-1]x_{\ell,i}^{2}.
\ese
For a large $\bF_{i}$, the covariance matrix $\bV_{i}$ is thus diagonally dominant. 
If $f_{\ell,i}$ is very large, then $C_{\ell,\ell,i}^{-1}\approx 1$ and $S_{\ell,\ell,i}\approx [\exp\{s^{2}(\|\bx_{i}\|_{2}/d)\}-1]x_{\ell,i}^{2}$. 
Due to the latter component, a large $f_{\ell,i}$ may not always ensure low conditional variability in $w_{\ell,i,j}$. 
If, however, smaller $x_{\ell,i}$'s correspond to larger values of $f_{\ell,i}$, then we have smaller conditional variability in $w_{\ell,i,j}$. 
As discussed in detail in \cite{sarkar2018bayesian}, by the very nature of such problems, the conditional variability of each component $w_{\ell,i,j}$ should depend primarily on the corresponding latent component $x_{\ell,i}$. 
Interestingly, however, unlike \cite{sarkar2018bayesian}, in our model the expected component-specific conditional variances $S_{\ell,\ell,i}$ of $w_{\ell,i,j}$ involve contributions from all entries of $\bx_{i}$.
This is not surprising as our characterization of the joint distribution of $\bw$ rely on two conditionally independent distributions, 
namely the distributions of its norm and its direction given the unobserved $\bx$. 
And the norm and direction of a vector are functions of all of its entries.
Primary dependence of $S_{\ell,\ell,i}$ on $x_{\ell,i}$ is, however, still accommodated via the functions $\kappa_{\ell}(x_{\ell,i})$ and the resulting diagonally dominant nature of $\bV_{i}$. 
\cite{Staudenmayer_etal:2008,sarkar2014bayesian,sarkar2018bayesian,sarkar2021bayesian} also showed that the variability in $w_{\ell,i,j}$ usually increases with $x_{\ell,i}$, especially in dietary recall data. 
It would have thus been ideal to have the functions $\kappa_{\ell}(x_{\ell,i})$'s be non-increasing in $x_{\ell,i}$ and the function $s^{2}(\|\bx\|_{2}/d)$ increasing in $\|\bx\|$.
Although we have not imposed such shape constraints explicitly in our model, the estimated functions do exhibit such behaviour in all our simulated and real data applications.
Even greater flexibility in the conditional covariance structure may be obtained by taking mixtures of MvMF distributions instead of one single MvMF. 
Such extensions will, however, be pursued elsewhere.

\subsection{Connections with Classical Models} \label{sec: classical}
Our proposed model can be rewritten as a classical additive measurement error model as $\bw_{i,j}=r_{i,j}\bC_{i}\bQ_{i,j}\bx_{i} = \bx_{i}+(r_{i,j}\bC_{i}\bQ_{i,j}-\bI_{d})\bx_{i} = \bx_{i}+\bu_{i,j}$ with $\bu_{i,j} = (r_{i,j}\bC_{i}\bQ_{i,j}-\bI_{d})\bx_{i}$ satisfying $\eE(\bu_{i,j}|\bx_{i})=\bzero$. 
The additive measurement error $\bu_{i,j}$ is thus the vector joining the tips of $\bx_{i}$ and $\bw_{i,j}$ (Figure \ref{fig: Stochastic Rotation}). 
However, unlike previous works on conditionally varying measurement errors such as \cite{sarkar2018bayesian, sarkar2021bayesian}, 
the entries of $\bw_{i,j}$ and hence those of $\bu_{i,j}$ are allowed to depend on all components of $\bx_{i}$. 

The proposed model also has a resemblance with classical multiplicative measurement error models 
$\bw_{i,j} = \bx_{i} ~\circ~ \wt\bu_{i,j}$
where $\circ$ denotes element-wise product 
and the errors $\wt\bu_{i,j}$ are distributed independently of $\bx_{i}$ with $E(\wt\bu_{i,j}) = \bone$. 
In our model, however, we have $\wt\bu_{i,j}=r_{i,j}\bC'_{i}\bQ_{i,j}\bx_{i}$, where $\bC'_{i}=\diag(1/\bx_{i})\bC_{i}$.

In both specifications, the scalar part $r_{i,j}$ is shared by all components of the measurement errors $\bu_{i,j}$ and $\wt\bu_{i,j}$. 
However, further interactions between the entries of $\bx_{i}$ and the entries of $\bu_{i,j}$ or $\wt\bu_{i,j}$ may be observed depending on the concentration of $\bQ_{i,j}$ around $\bI_{d}$. 
For example, when $\bQ_{i,j}=\bI_d$ for all $j$, there is only an element-wise effect of $\bx_{i}$'s on the errors. 
However, as $\bQ_{i,j}$ moves away from $\bI_{d}$, there is more inter-component associations between the $\bx_{i}$'s and the associated errors $\bu_{i,j}$ or $\wt\bu_{i,j}$.

In the following subsection, 
we show that a classical additive normal measurement error model can be obtained as a meaningful limiting case of our rotation-based model.

\vskip-5pt
\subsubsection{Limiting Equivalence with a Normal Error Model}

To reduce notation complexity, we initially fix $r_{i,j}=1$ for all $i,j$ in this section.
The MGF of $\bQ_{i,j}$ is $\eE\{\etr(\bQ_{i,j}\bT)\}=\frac{M(\bF_{i}+\bT)}{M(\bF_{i})}$.
We recall that $\bF_{i}=\diag(\bFf_{i})$ 
and we have argued that in practice its entries are expected to be large.
We now study the limiting distribution of each coordinate of $\bu'_{i,j}=\bH_{i}^{1/2}(\bC_{i}\bQ_{i,j}-\bI_{d})\bx_{i}$, where $\bH_{i}=\diag(\bh_i)$ with $h_{\ell,i}=x_{\ell,i}^{-2}\left\{1-\sum_{k\neq \ell}\frac{1}{2(f_{\ell,i}+f_{k,i})}\right\}^2\left\{\sum_{k\neq \ell}\frac{1}{2(f_{\ell,i}+f_{k,i})^2}\right\}^{-1}$. It is easy to see that $h_{\ell,i}\rightarrow\infty$ as all the entries in $\bFf_{i}\rightarrow\infty$. 

Without any loss of generality, we set $\bt=(t_{1},0,\ldots,0)$.
We have $\eE\{\exp(\bt\trans\bH_{i}^{1/2}\bC_{i}\bQ_{i,j}\bx_{i})\}=\eE\{\etr(\bQ_{i,j}\bx_{i}\bt\trans\bH_{i}^{1/2}\bC_{i})\}=\frac{M(\bF_{i}+\bx_{i}\bt\trans\bH_{i}^{1/2}\bC_{i})}{M(\bF_{i})}$.
When $\bF_{i}$ is large, $\bF_{i}+\bx_{i}\bt\trans\bH_{i}^{1/2}\bC_{i}$ will be diagonally dominant and its off-diagonal entries will be very small.
Hence, the eigenvalues of $\bF_{i}+\bx_{i}\bt\trans\bH_{i}^{1/2}\bC_{i}$ will be its diagonal entries.
Applying the approximation from \eqref{eq:normconst}, we then have 
\bse
    &&\frac{M(\bF_{i}+\bx_{i}\bt\trans\bH_{i}^{1/2}\bC_{i})}{M(\bF_{i})}\approx \exp(h_{1,i}^{1/2}C_{1,1,i}x_{1,i}t_{1})\left\{\frac{\prod_{k\neq 1}(f_{1,i}+f_{k,i})}{\prod_{k\neq 1}(f_{1,i}+f_{k,i}+h_{1,i}^{1/2}C_{1,1,i}x_{1,i}t_{1})}\right\}^{1/2}\\
    &&\quad = \exp(h_{1,i}^{1/2}C_{1,1,i}x_{1,i}t_{1})\exp\left\{-\frac{1}{2}\sum_{k\neq 1}\log\left(1+\frac{h_{1,i}^{1/2}C_{1,1,i}x_{1,i}t_{1}}{f_{1,i}+f_{k,i}}\right)\right\}\\
    &&\quad\approx \exp(h_{1,i}^{1/2}C_{1,1,i}x_{1,i}t_{1})\exp\left\{-\sum_{k\neq 1}\frac{1}{2}\frac{h_{1,i}^{1/2}C_{1,1,i}x_{1,i}t_{1}}{f_{1,i}+f_{k,i}}+\sum_{k\neq 1}\frac{1}{4}\left(\frac{h_{1,i}^{1/2}C_{1,1,i}x_{1,i}t_{1}}{f_{1,i}+f_{k,i}}\right)^2 + O\left(\max_k\frac{1}{f_{k,i}^{3/2}}\right)\right\},
\ese
applying Taylor series expansion.
Since $C_{1,1,i} \approx 1\bigg/\left\{1-\sum_{k\neq 1}\frac{1}{2(f_{1,i}+f_{k,i})}\right\}$, we can simplify the above expression further to $\exp(h_{1,i}^{1/2}x_{1,i}t_{1})\exp\left\{\frac{1}{4}\sum_{k\neq 1}\left(\frac{h_{1,i}^{1/2}C_{1,1,i}x_{1,i}t_{1}}{f_{1,i}+f_{k,i}}\right)^2\right\}$ 
which reduces to 
\begin{align*}
\exp(h_{1,i}^{1/2}x_{1, i}t_1)\exp\left\{\frac{1}{2}h_{1,i}C_{1,1,i}^2x_{1, i}^2t_1^2\sum_{k\neq 1}\frac{1}{2(f_{1,i}+f_{k,i})^2}\right\} 
= \exp(h_{1,i}^{1/2}x_{1, i}t_1)\exp(t_{1}^{2}/2).
\end{align*}
Hence, each coordinate of $\bu'_{i,j}$ marginally approaches to $\Normal(0,1)$. 
Thus, in a limiting sense, the $\ell\th$ coordinate of $\frac{\bw_{i,j}}{r_{i,j}}$ marginally reduces to a normally distributed classical measurement error model with mean $x_{\ell, i}$ and variance $h_{\ell,i}^{-1}$.

\subsection{Error Distribution Generating Function}
To characterize the distribution of $\bu_{i,j}$, we can compute the moment generating function (MGF) of $(\bu_{i,j}\vert\bx_{i})$ using the MGF result in equation (2.7) of \cite{khatri1977mises} as
\bse
   && \eE\{\exp(\bt\trans\bu_{i,j}) \vert \bx_{i}\}=\exp(-\bt\trans\bx_{i})\eE[\exp\{\trace(r_{i,j}\bx_{i}\bt\trans\bC_{i}\bQ_{i,j})\}\vert \bx_{i}] \nonumber\\
   && =\exp(-\bt\trans\bx_{i})\eE\left\{\frac{M(\bF_{i}+r_{i,j}\bx_{i}\bt\trans\bC_{i})}{M(\bF_{i})} \vert \bx_{i}\right\} = \exp(-\bt\trans\bx_{i})\frac{1}{M(\bF_{i})}\eE\left\{M(\bF_{i}+r_{i,j}\bx_{i}\bt\trans\bC_{i}) \vert \bx_{i}\right\}, 
\ese
where $M(\cdot)$ is the normalizing constant whose approximation is given in~\eqref{eq:normconst} for large $\bF_{i}$. 
To study the tail properties of each component in $\bu_{i,j}$, we can consider different choices of $\bt$. 
Specifically, for studying the $\ell\th$ component, the choice is $\bt_{\ell}=(0,\dots,t_{\ell},\dots,0)\trans$. 
Assuming $\bF_{i}$ is large, the approximation in~\eqref{eq:normconst} gives us 
\bse
& \exp(-\bt\trans\bx_{i})\frac{1}{M(\bF_{i})}\eE\left\{M(\bF_{i}+r_{i,j}\bx_{i}\bt\trans\bC_{i}) \vert \bx_{i}\right\}\\ 
& \approx\exp(-\bt\trans\bx_{i})\eE\left[\exp(r_{i,j}\bt\trans\bC_{i}\bx_{i})\prod_{l=2}^d\left\{1+\frac{r_{i,j}\bt\trans\bC_{i}\bx_{i}}{\kappa_{l}(x_{l,i})+\kappa_{1}(x_{1,i})}\right\}^{-\frac{1}{2}}\vert\bx_{i}\right]. 
\ese
Explicit characterization of the MGF beyond the above expression is difficult to obtain.
We see, however, that the MGF might be undefined for positive valued $\bt$ for the heavy-tailed log-normal distribution on the $r_{i,j}$'s.
Other light-tailed choices, 
such as a gamma distribution with the rate parameter modeled as a function of $\|\bx_{i}\|_{2}$, may also be considered.
In this article, however, we focus on the log-normal.

\vskip 45pt
\subsection{Bayesian Inference} \label{sec: bayes}
While the main idea underlying our rotation-based approach to multivariate deconvolution is clearly statistical paradigm generic, 
in this article, we adopt a Bayesian route to estimation and inference. 
The rest of this section discusses prior specification, posterior computation and posterior convergence in such settings.

\subsubsection{Prior Specification} \label{sec: prior}
Since the functions $s^{2}(\cdot)$ and $\kappa_{\ell}(\cdot)$'s are strictly positive, we put truncated normal priors, truncated to $[0,\infty)$, on the associated B-spline coefficients: $\beta_{s,k}\sim \TN_{[ 0, \infty]}(\mu_{s,k},\sigma_{s}^{2})$ and $\beta_{\kappa,\ell,k}\sim\TN_{[ 0, \infty]}(\mu_{\kappa,\ell,k},\sigma_{\kappa,\ell}^{2})$. 
We assign Gamma priors on the inverse variances  $\sigma_{s}^{-2}\sim\Ga(a_{s}, b_{s})$ and $\sigma_{\kappa,\ell}^{-2}\sim \Ga(a_{\kappa,\ell}, b_{\kappa,\ell})$. 
For the component specific parameters $(\mu_{k},\sigma_{k}^{2})$ of the truncated normal mixtures characterizing the marginals $f_{x_{\ell}}$, we assign semi-conjugate independent priors $\mu_{k} \sim \Normal(\mu_{0},\sigma_{0}^{2})$ and $\sigma_{k}^{-2} \sim \Ga(a_{0}, b_{0})$. 
We assign independent Dirichlet priors on the mixture probabilities $\bpi_{\ell} = (\pi_{\ell,1},\dots,\pi_{\ell,K})\trans \sim \Dir(\alpha/K,\dots,\alpha/K)$, 
and independent uniform priors on the polar angles $\zeta_{m,j} \sim \Unif[0,\pi]$ and $\zeta_{m,m-1} \sim \Unif[0,2\pi]$.
The choice for the hyper-parameters are $a_{s}=b_{s}=a_{\kappa,\ell}=b_{\kappa,\ell}=0.1$. 
The hyper-parameter choices for $\mu_{s,k}$ and $\mu_{\kappa,\ell,k}$ are obtained based on the estimated $\bx_{i}$'s from the univariate sampler. 
The hyper-parameters $\mu_{0}$ and $\sigma_{0}$ from the copula model of $\bx$ are also set based on these univariate estimates.
The other two hyper-parameters $a_0$ and $b_0$ are set as $a_0=b_0=1$. 
The univariate sampler follows the additive model of \cite{sarkar2021bayesian}. 
Details are provided in Section~\ref{sec: post comp}.

\subsubsection{Posterior Computation} \label{sec: post comp}

Our inference is based on samples drawn from the posterior using an MCMC algorithm. 
The joint log posterior distribution of the model parameters is given by 
\begin{align*}
    &\sum_{i,j}\left[-\log\{M(\bF_{i})\}+\left\{{\mathcal{G}\left(\bC_{i}^{-1}\bw_{i,j},\bx_{i}\right) \bF_{i}}\right\}\right]-\frac{1}{2}\sum_{i,j}\log\{\|\bC_{i}^{-1}\bw_{i,j}\|^2_{2}/\|\bx_{i}\|^2_{2}s^{2}(\|\bx_{i}\|_{2}/d)\}\\
    &-\frac{1}{2}\sum_{i,j}\{\log(\|\bC_{i}^{-1}\bw_{i,j}\|_{2}/\|\bx_{i}\|_{2})+s^{2}(\|\bx_{i}\|_{2}/d)/2\}^{2}/s^{2}(\|\bx_{i}\|_{2}/d)-\sum_{i}\log f_{\bx}(\bx_{i})\\
     &
     -\frac{1}{2}\sum_{\ell,k} (\beta_{\kappa,\ell,k}-\mu_{\kappa,\ell,k})^{2}\mathbf{1}_{\beta_{\kappa,\ell}\in[0,\infty]}/\sigma_{\kappa,\ell}^2-\frac{1}{2}\sum_{k}(\beta_{s,k}-\mu_{s,k})^2\mathbf{1}_{\beta_{s,k}\in[0,\infty]}/\sigma_{s}^2\\
     & -\frac{1}{2}\sum_{k}(\mu_{k}-\mu_{0})^{2}/\sigma_{0}^{2} -\sum_{k}(a_{0}+1)\log(\sigma_{k}^{2})-\sum_{k} b_{0}/\sigma_{k}^{2}.
\end{align*}

It is possible to efficiently calculate the derivatives of the above likelihood with respect to $\bbeta_{s}$, $\mu_{\ell,k}$, $\sigma_{\ell,k}$, etc. 
These parameters may therefore be updated using HMC algorithms \citep{neal2011mcmc, betancourt2015hamiltonian, betancourt2017conceptual}. 
HMC has been shown to draw posterior samples much more efficiently than traditional random walk Metropolis-Hastings in complex Bayesian hierarchical models \citep{betancourt2015hamiltonian} by more efficiently exploring the target distribution under local correlations among the parameters.
A short review of HMC sampling is provided in Section S.2 in the supplementary materials for easy reference.
The $\bx_{i}$'s, and the parameters specifying the $\bR_{\bx}$ of $f_{\bx}(\bx_{i})$, can be updated using random walk Metropolis-Hastings steps.  
The parameters $\bbeta_{\kappa}$ are updated using adaptive Metropolis-Hastings.
Other parameters have closed-form full conditionals and can be straightforwardly updated. 
Details are deferred to Section S.5 in the supplementary materials.

\section{Simulation Experiments} \label{sec: sim studies}
In this section, we evaluate the performance of our proposed method, referred to in the tabular result summaries below as the DeStoR method, short for `Deconvolution via Stochastic Rotation'.
We compare with the method of \cite{sarkar2021bayesian}, referred to as the SPMC method here following the first letters of the authors' last names, 
which was shown to vastly outperform the only two previously existing multivariate deconvolution methods accommodating heteroscedastic measurement errors, 
namely the pseudo-Bayesian two-stage method of \cite{Zhang2011b} and the multivariate mixture normal based method of \cite{sarkar2018bayesian}. 
We consider two different simulation scenarios - (a) a well-specified case; and (b) a mis-specified case. 
For both scenarios, we consider the same $f_{\bx}$ to generate the true latent $\bx_{i}$'s 
but use different choices for $f_{\bw \vert \bx}$ to generate the replicates $\bw_{i,j}$'s. 
Our choice for the well-specified case conforms to our the proposed formulation for $f_{\bw \vert \bx}$ in \eqref{eq:Orthmoddel} in Section \ref{sec: models}. 
The mis-specified case is designed to evaluate the robustness of the proposed method to deviations from model assumptions and considers additive measurement errors generated from a Gaussian copula model with a mean restricted mixture of normals for the marginals from \cite{sarkar2021bayesian} to produce the replicates $\bw_{i,j}$'s.  

While our proposed method scales well to much higher dimensional problems, 
we consider a relatively low $d=3$ dimensional problem here as 
the computation of the joint density on a $3$ dimensional grid remains manageable 
and the results for ${{3}\choose{2}} = 3$ bivariate marginals can also be conveniently graphically summarized. 
We generate the true $x_{\ell,i}$'s for $\ell=1,\dots,d$ as follows. We (a) first sample $\bx_{i}^{\triangle} \sim \MVN_{d}(\bzero,\bR_{\bx})$, (b) then, set $\bx_{i}^{\triangle\triangle}=\Phi(\bx_{i}^{\triangle})$, (c) finally, set $x_{\ell,i} = F_{TN,mix}^{-1}(x_{\ell,i}^{\triangle\triangle} \vert \bpi_{x,\ell},\bmu_{x,\ell},\bsigma_{x,\ell}^{2}, A,B)$, where $F_{TN,mix}(X \vert \bpi,\bmu,\bsigma^{2},x_{L},x_{U}) = \sum_{k=1}^{K}\pi_{k}F_{TN}(X \vert \mu_{k},\sigma_{k}^{2},x_{L},x_{U})$. This way, the marginal distributions are mixtures of truncated normal distributions and hence can take widely varying shapes while the correlation between different components is $\bR_{\bx}$. 
We set 
\bse
\bR_{\bx} = \left(\begin{array}{c c c}

1 & 0.7 & 0.7^{2} \\

 & 1 & 0.7 \\

 &  & 1 

\end{array} \right),\bpi_{x,\ell} =  \left(\begin{array}{c}

0.25 \\

0.50 \\

0.25 

\end{array}\right) ~\text{for all}~\ell,
~~\bmu_{\bx} =  \left(\begin{array}{c}

\bmu_{x,1}^{\trans} \\

\bmu_{x,2}^{\trans} \\

\bmu_{x,3}^{\trans}

\end{array} \right) = \left(\begin{array}{c c c}

2 & 2 & 3 \\

2 & 3 & 5 \\

2 & 2 & 5 

\end{array} \right),
\ese
and $A=0, B=6,\sigma_{x,\ell,k}^{2}=0.75^{2}~\text{for all}~\ell,k$.

To generate $\bw_{i,j}$ for the well-specified case based on \eqref{eq:Orthmoddel}, we need to generate $\bQ_{i,j}$ and $r_{i,j}$ given $\bx_{i}$ that are generated in the previous step. 
We use the R package {\tt rstiefel} \citep{stiefelpkg} to generate $\bQ_{i,j}$ with concentration parameters $\kappa_{\ell}(x_{\ell,i})$'s where  $\kappa_{\ell}(x_{\ell,i})=60/x_{\ell,i}$. 
The scalars $r_{i,j}$'s are generated from log-normal distribution with mean = $s^{2}(\|\bx_{i}\|_{2}/d)/2$ and variance = $s^{2}(\|\bx_{i}\|_{2}/d)$, where $s(\|\bx_{i}\|_{2}/d)=\|\bx_{i}\|_{2}/150$.

While generating $\bw_{i,j}$ for the mis-specified case based on \cite{sarkar2021bayesian}, we consider the additive model $\bw_{i,j}=\bx_{i}+s_{1}(\bx_{i})\circ\bepsilon_{i,j}$, where $s_{1}(\bx_{i})=\bx_{i}/4$ and the  $\bepsilon_{i,j}$'s are generated from a standard multivariate normal distribution.

The integrated squared error (ISE) of estimation of $f_{\bx}$ by $\wh{f}_{\bx}$ is defined as $ISE = \int \{f_{\bx}(\bx)-\widehat{f}_{\bx}(\bx)\}^{2}d\bx$.
A Monte Carlo estimate of ISE is given by
$ISE_{est} = \sum_{m=1}^{M}\{f_{\bx}(\bx_{m})-\widehat{f}_{\bx}(\bx_{m})\}^{2}/p_{0}(\bx_{m})$,
where $\{\bx_{m}\}_{m=1}^{M}$ are random samples from the density $p_{0}$. 
We used the true densities $f_{\bx}$ for $p_{0}$ and the true values of the $\bx_{i}$'s for the $\bx_{m}$'s. 
For the univariate marginals, a Monte Carlo estimate of ISE is given by
$\sum_{i=1}^{N}\{f_{x}(x_{i}^{\Delta})-\wh{f}_{x}(x_{i}^{\Delta})\}^{2}  \Delta_{i}$,
where $\{x_{i}^{\Delta}\}_{i=0}^{N}$ are a set of grid points on the range of $x$ and
$\Delta_{i} = (x_{i}^{\Delta} - x_{i-1}^{\Delta})$ for all $i$.

\begin{table}[htbp]
    \centering
    \caption{Median integrated squared errors (MISE) of estimating $f_{\bx}$ by our method (DeStoR) and the method of \cite{sarkar2021bayesian} 
    when the replicates $\bw_{i,j}$ are generated from (a) our model (well-specified case), and (b) the model of \cite{sarkar2021bayesian} (mis-specified case).}
    \begin{tabular}{|c|rrr|r|}
    \hline
    \multicolumn{5}{|c|}{MISE $\times 1000$}\\
   \hline
    Method & Comp 1& Comp 2&Comp 3&3D-joint\\
    \hline
    \hline
    & \multicolumn{4}{c|}{Well Specified Case}\\
    \hline
    DeStoR  &  0.10 &0.39& 2.12&0.96\\
     SPMC & 0.95 &1.68 &3.08 &3.79\\
    \hline
    & \multicolumn{4}{c|}{Mis-specified Case}\\
    \hline
     DeStoR  &  1.77 &1.62& 0.68&2.95\\
     SPMC & 0.94 &3.55 &2.71&1.14\\
     \hline
    \end{tabular}
    \label{tab: MISEs}
\end{table}

Table \ref{tab: MISEs} reports the median ISEs (MISEs) for estimating the trivariate joint densities and the univariate marginals obtained by our method and the method of \cite{sarkar2021bayesian}. 
The reported MISEs are all based on $100$ simulated data sets.
In the well-specified case, when the data-generating mechanism conforms to our proposed model, our method significantly outperformed \cite{sarkar2021bayesian} in estimating the three-dimensional joint density and as well as all univariate marginals. 
In the mis-specified case, when the data generating mechanism conforms to the model of \cite{sarkar2021bayesian}, our method still performed competitively with \cite{sarkar2021bayesian} for the three-dimensional joint density estimation problem and actually outperformed \cite{sarkar2021bayesian} for some of the univariate marginal density estimation problems. 
We attribute this to our more efficient MCMC sampling of the posterior via HMC samplers.

\begin{figure}
    \centering
    \includegraphics[width = 1\textwidth]{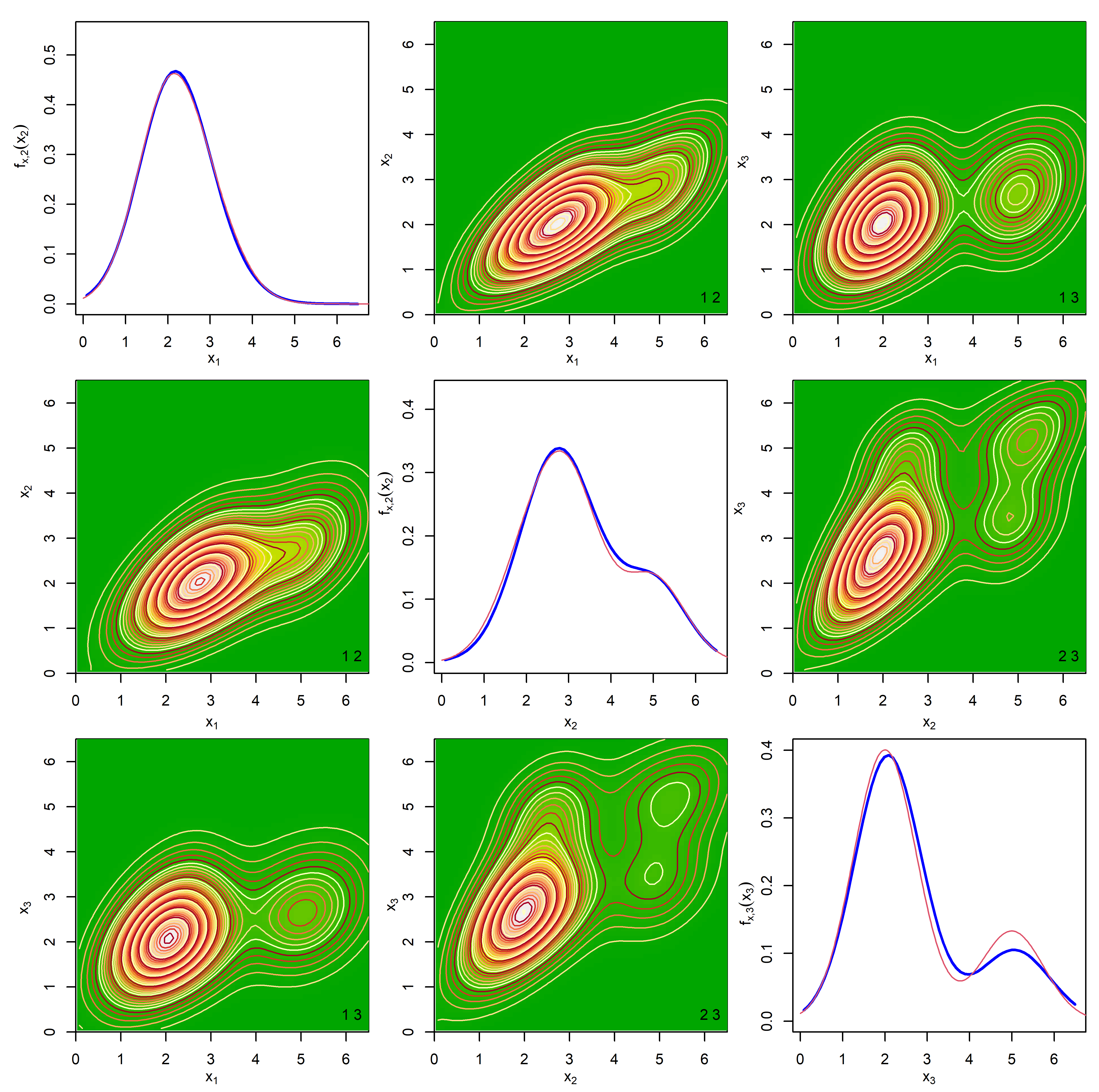}
    \caption{Results for the simulated data with $n=1000$ subjects and $m_{i}=3$ replicates per subject when true data generating process follows the structure  \eqref{eq:Orthmoddel} proposed in this article. The figures in the diagonal panels illustrate the true marginal densities $f_{\bx,\ell}(x_{\ell})$ in red and the corresponding estimates produced by our method in blue. The figures in the off-diagonal panels depict the contour plots of the true two-dimensional marginals (upper triangular panels) and the corresponding estimates obtained by our method (lower triangular panels).}
    \label{fig:simplot21}
\end{figure}

\begin{figure}
    \centering
    \includegraphics[width = 1\textwidth]{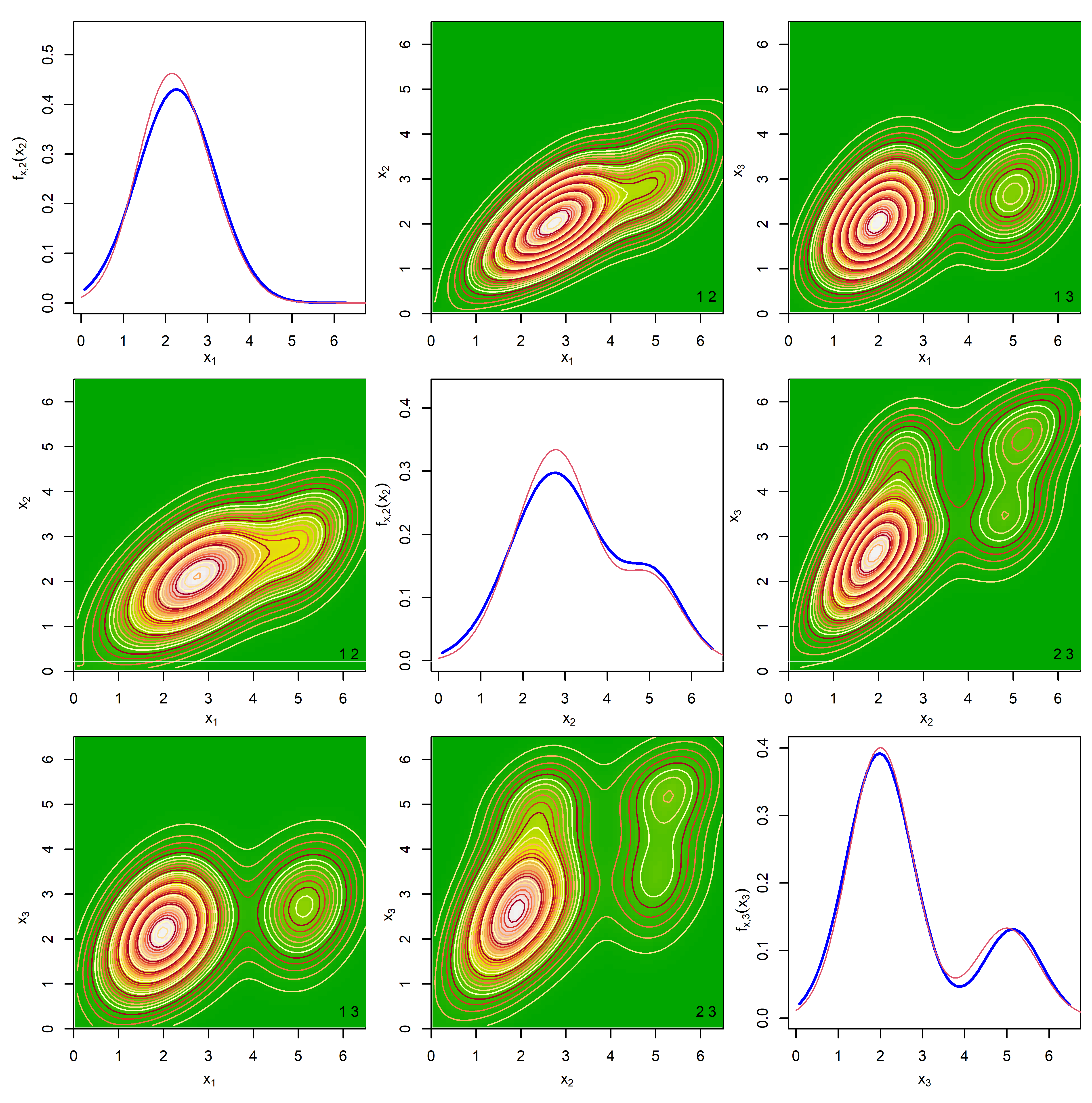}
    \caption{Results for the simulated data with $n=1000$ subjects and $m_{i}=3$ replicates per subject when true data generating process follows the additive model from \cite{sarkar2021bayesian}. The figures in the diagonal panels illustrate the true marginal densities $f_{\bx,\ell}(x_{\ell})$ in red and the corresponding estimates produced by our method in blue. The figures in the off-diagonal panels depict the contour plots of the true two-dimensional marginals (upper triangular panels) and the corresponding estimates obtained by our method (lower triangular panels).}
    \label{fig:simplot22}
\end{figure}

Figures \ref{fig:simplot21} and \ref{fig:simplot22} 
show the estimates of the univariate marginal and bivariate joint densities obtained by our method for the data set that produced the $25$ percentile ISE and the corresponding truths for the well-specified and the mis-specified cases, respectively. 
The estimates clearly provide excellent fits to the truths. 
Additional plots comparing the estimates of the univariate marginals produced by our method with those obtained by the method of \cite{sarkar2021bayesian} are presented in Figures S.3 and S.4 in Section S.9 of the supplementary materials.  

We also evaluate the numerical performance of the proposed method in some higher dimensional cases, namely $d=5$ and $d=10$. 
We maintain a similar structure for $\bR_{\bx}$ and set $\sigma^{2}_{x,\ell,k}=0.75^{2}$ for all the univariate components as before.
The mixture probabilities are also kept the same, as the univariate distributions are again assumed to be mixtures of three univariate normals. 
However, since it is difficult to set the mean parameter $\bmu_{\bx}$ in higher dimension explicitly, we set the components of $\bmu_{\bx}$ using random numbers generated from $\Unif(1, 10)$. 
We summarize the results in Table \ref{tab: MISEs 2}. 
For clarity of our presentation, here we only present the MISEs for the $d$-dimensional joint densities. 
The results show patterns similar to that previously observed for the $d=3$ case.

\begin{table}[htbp]
    \centering
    \caption{Median integrated squared errors (MISE) of estimating $f_{\bx}$ by our method (DeStoR) and the method of \cite{sarkar2021bayesian} 
    when the replicates $\bw_{i,j}$ are generated from (a) our model (well-specified case), and (b) the model of \cite{sarkar2021bayesian} (mis-specified case).}
    \begin{tabular}{|c|r|r|}
    \hline
    \multicolumn{3}{|c|}{Joint MISE $\times 1000$}\\
   \hline
    Method & $d=5$ & $d=10$\\
    \hline
    \hline
    & \multicolumn{2}{c|}{Well-specified Case}\\
    \hline
    DeStoR   & 1.72&1.62\\
     SPMC & 2.58 &2.79\\
    \hline
    & \multicolumn{2}{c|}{Mis-specified Case}\\
    \hline
     DeStoR  & 3.68&3.91\\
     SPMC &2.71&2.54\\
     \hline
    \end{tabular}
    \label{tab: MISEs 2}
\end{table}

\section{Real Data Application} \label{sec: applications}

The estimation of the joint and marginal distributions of long-term average daily intakes of different dietary components is a fundamentally important
problem in nutritional epidemiology.
The long-term average daily intakes of the dietary components, $\bx$, can not, however, be directly measured.
Data are thus often collected via nutritional surveys in the form of dietary recalls, 
the subjects participating in the study remembering and reporting the type and amount of food they consumed in the past 24 hours.
The problem of estimating the joint consumption pattern of the dietary components from the contaminated 24-hour recalls then becomes a problem of multivariate density deconvolution.  

One such large-scale nutritional survey is 
the Eating at America's Table (EATS) study \citep{Subar2001} conducted by the National Cancer Institute where $n=965$ participants were interviewed $m_{i}=4$ times over the course of a year 
and their 24 hour dietary recalls $\bw_{i,j}$ were recorded.
The goal is to estimate the joint and marginal consumption patterns of the underlying true daily intakes $\bx_{i}$.
In this article, we focus particularly on the average daily intakes of three minerals, namely, iron, magnesium, and sodium.

\begin{figure}[!ht]
    \centering
    \includegraphics[width = 0.95\textwidth, height=0.75\textheight]{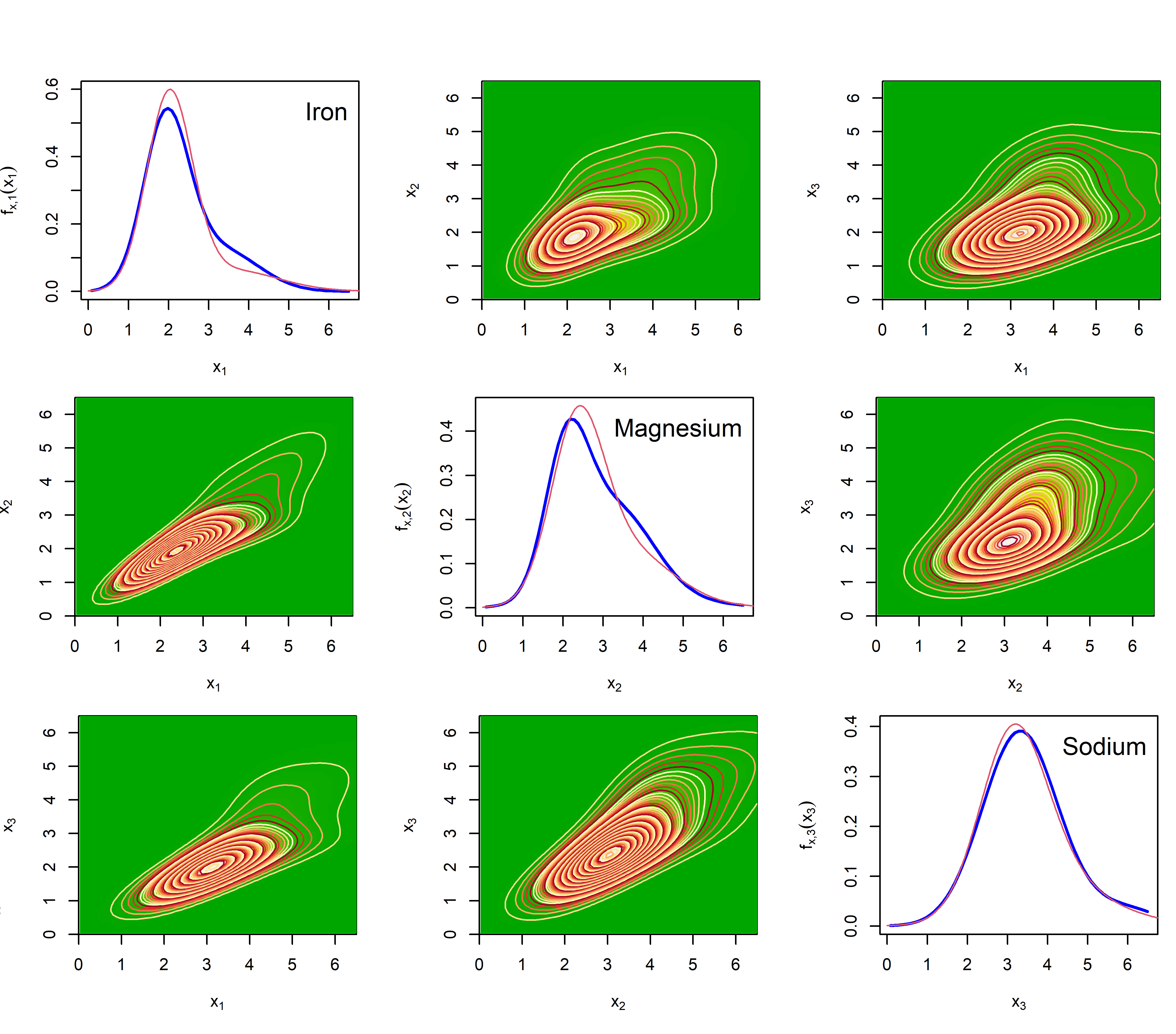}
    \caption{Results for the EATS data with $n=965$ subjects and $m_{i}=4$ recalls per subject for three minerals, namely, $x_{1}=$ iron, $x_{2}=$ magnesium, and $x_{3}=$ sodium. The figures in the diagonal panels illustrate the estimated marginal densities $f_{\bx,\ell}(x_{\ell})$, obtained from our method in blue and the method of \cite{sarkar2021bayesian} in red. The figures in the off-diagonal panels depict the contour plots of the estimated two-dimensional marginals 
    obtained from our method (upper triangular panels) and the method of \cite{sarkar2021bayesian} (lower triangular panels).}
    \label{fig:realplot2}
\end{figure}

Figure~\ref{fig:realplot2} presents the estimated univariate and bivariate marginals obtained by our proposed method and the method of \cite{sarkar2021bayesian}. 
The estimates of the univariate marginals produced by the two methods are quite similar.
The estimated bivariate densities, however, although share some commonalities across the two methods, are not exactly the same. 
Specifically, the estimates produced by the method of \cite{sarkar2021bayesian} are more strongly correlated than ours. 

We conducted a formal model comparison exercise to identify which model fits the EATS data set better. 
Since the competing methods are both Bayesian, 
we compared them in terms of the Bayes factor \citep{kass1995bayes}. 
The Bayes factor comparing the two methods is given by
$B=\frac{P(D^{(i)}|\textrm{DeStoR})}{P(D^{(i)}|\textrm{SPMC})}$.
We calculated the numerator and denominator from the posterior samples using the harmonic mean identity of \cite{neton1994approximate}.
Based on the suggestions in \cite{kass1995bayes}, the evidence in favor of our proposed DeStoR is `decisive' as we obtained $B>100$.

\section{Discussion} \label{sec: discussion}
In this article, we developed a novel method for multivariate density deconvolution in the presence of conditionally heteroscedastic measurement errors and the availability of replicated proxies for the unknown values of the variable of interest. 
Our proposed method relies on stochastically rotating the observed replicates toward the underlying true values and then stochastically adjusting their lengths to match the lengths of the true values. 
We took a Bayesian route to estimation and inference, implemented via an efficient MCMC algorithm. 
In synthetic numerical experiments, the proposed method showed excellent performance in recovering the true density of interest. 
The method's practical utility was demonstrated in a nutritional epidemiology application in estimating the joint distribution of the true average long-term intakes of three different dietary components from their measurement error-contaminated 24-hour recalls.

{In Section S.6 of the supplementary materials, 
we show that the posterior of our model convergences to the true unknown density of the observed data $\bw_{i}$ as the sample size grows to infinity. 
There is substantial literature studying the convergence properties of the posteriors in ordinary density estimation problems where $f_{\bw}$ is directly modeled. 
Our results are, however, established under a conditionally heteroscedastic measurement error setting as described in (\ref{eq:Orthmoddel}), 
where $f_{\bw}$ is obtained 
by a convolution of the models for $f_{\bx}$ and $f_{\bu \mid \bx}$. 
To our knowledge, our results are novel to the literature, especially under dependence between the measurement error $\bu$ and the unknown true vector of interest $\bx$. 
In the context of density deconvolution, however, our results lack strong inferential merit as the notion of recovery is in terms of $f_{\bw}$ but not $f_{\bx}$.
In the future, we hope to study posterior consistency with a more appropriate notion of recovery, that of the density of interest $f_{\bx}$, under density deconvolution with conditional heteroscedastic matrix vMF distributed measurement errors.}

While this is our first work in this line of research, our proposed approach is already quite general, not only allowing completely unknown measurement error distributions but also accommodating unknown conditional heteroscedasticity patterns. 
We thus believe the work contributes a novel and significant addition to the existing sparse literature on multivariate density deconvolution.

Directions for methodological extensions and topics of our ongoing research include exploration of other distributions on the Stiefel manifold, including mixtures of MvMF, 
adaptations to regression problems with errors-in-covariates, 
etc.

\baselineskip=17pt

\section*{Supplementary Material}
The supplementary material presents brief reviews of 
copulas, 
and Hamiltonian Monte Carlo, 
and also the explicit formula of cubic B-splines 
for easy reference;
discusses model identifiability; 
and details the choice of hyper-parameters, 
the MCMC algorithm used to sample from the posterior, 
and some convergence results for the posterior and their proofs.  
The supplementary material also presents some additional figures summarizing the results of the simulation experiments. 
R programs implementing the deconvolution methods developed in this article are included in the supplementary material. 
The EATS data analyzed in Section \ref{sec: applications} can be accessed from National Cancer Institute by arranging a Material Transfer Agreement.
A simulated data set, generated using the estimates produced by our method for the EATS data set, 
and a `readme' file providing additional details are also included in the supplementary material.

\section*{Acknowledgements}
We thank two anonymous referees for constructive suggestions that led to significant improvement in the exposition of the manuscript. 

\bibliographystyle{natbib}
\bibliography{main}

\clearpage\pagebreak\newpage
\pagestyle{fancy}
\fancyhf{}
\rhead{\bfseries\thepage}
\lhead{\bfseries SUPPLEMENTARY MATERIAL}

\begin{center}
\baselineskip=27pt
{\LARGE Supplementary Material for\\ 
{\bf Bayesian Semiparametric\\ Multivariate 
Density Deconvolution via 
Stochastic Rotation of Replicates
}}
\end{center}

\vskip 2mm
\begin{center}
Arkaprava Roy\\
ark007@ufl.edu\\
Department of Biostatistics,
University of Florida\\
2004 Mowry Road, Gainesville, FL  32611, USA\\
\vskip 2mm
Abhra Sarkar\\
abhra.sarkar@utexas.edu \\
Department of Statistics and Data Sciences,
The University of Texas at Austin\\
2317 Speedway D9800, Austin, TX 78712-1823, USA\\
\end{center}

\setcounter{equation}{0}
\setcounter{page}{1}
\setcounter{table}{1}
\setcounter{figure}{0}
\setcounter{section}{0}
\numberwithin{table}{section}
\renewcommand{\theequation}{S.\arabic{equation}}
\renewcommand{\thesubsection}{S.\arabic{section}.\arabic{subsection}}
\renewcommand{\thesection}{S.\arabic{section}}
\renewcommand{\thepage}{S.\arabic{page}}
\renewcommand{\thetable}{S.\arabic{table}}
\renewcommand{\thefigure}{S.\arabic{figure}}
\baselineskip=16pt

\vskip 10mm
Supplementary material presents 
brief reviews of 
copula basics, 
Hamiltonian Monte Carlo, 
and cubic B-splines to make the article relatively self-contained. 
Supplementary material also 
discusses model identifiability;
and presents 
details of the MCMC algorithm we designed to sample from the posterior; 
some convergence results for the posterior and their proofs; 
and some additional figures. 
Separate files additionally include a synthetic data set, generated using the estimates produced by our method for the EATS data analyzed in Section 5 of the main paper, 
and \texttt{R} programs implementing the proposed multivariate density deconvolution model developed in this article.

\newpage 
\section{Gaussian Copula} \label{sec: copula basics}
The literature on copula models is enormous. See, for example, \citelatex{nelsen2007introduction, joe2015dependence, shemyakin2017introduction} and the references therein.
For easy reference, we provide a brief review of the basics here. 

A function $\cC(\bu) = \cC (u_{1},\dots,u_{p}) : [0,1]^p \rightarrow [0,1]$ is called a copula 
if $\cC(\bu)$ is a continuous cumulative distribution function (cdf) on $[0,1]^p$ 
such that each marginal is a uniform cdf on $[0,1]$. 
That is, for any $\bu \in [0,1]^p$,
$\cC(\bu) = \cC (u_{1},\dots,u_{p}) = \Pr(U_{1}\leq u_{1}, \dots,U_{p}\leq u_{p})$ with 
$\cC(1,\dots,1,u_{i},1,\dots,1) = \Pr(U_{i} \leq u_{i}) = u_{i},  i=1,\dots,p$.
If $\{X_i\}_{i=1}^{p}$ are absolutely continuous random variables having marginal cdf $\{H_{i}(x_{i})\}_{i=1}^{p}$ 
and marginal probability density functions (pdf) $\{h_{i}(x_{i})\}_{i=1}^{p}$,  joint cdf $H(x_{1},\dots,x_{p})$ and joint pdf $h(x_{1},\dots,x_{p})$, 
then a copula $\cC$ can be defined in terms of $H$ as 
$\cC(u_{1},\dots,u_{p}) = H \left(x_{1}, \dots, x_{p}\right)$ where $u_{i} = H_{i}(x_{i}), i=1,\dots,p$.
It follows that
$h(x_{1},\dots,x_{p}) = c(u_{1},\dots,u_{p})\prod_{i=1}^{p}h_{i}(x_{i})$, 
where $c(u_{1},\dots,u_{p}) = {\partial^p \cC(u_{1},\dots,u_{p})}  /  {(\partial u_{1} \dots \partial u_{p})}$.
This defines a copula density $c(\bu)$ in terms of the joint and marginal pdfs of $\{X_{i}\}_{i=1}^{p}$ as
\vspace{-4ex}\\
\be
\textstyle c(u_{1},\dots,u_{p}) = h(x_{1},\dots,x_{p}) / \prod_{i=1}^{p}h_{i}(x_{i}). \label{eq:Copula A2}
\ee
\vspace{-4ex}\\
Conversely, if $\{V_i\}_{i=1}^{p}$ are continuous random variables having fixed marginal cdfs $\{F_{i}(v_i)\}_{i=1}^{p}$, 
then their joint cdf $F(v_{1},\dots,v_{p})$, with a dependence structure introduced through a copula $\cC$, can be defined as
\vspace{-4ex}\\
\be
F(v_{1},\dots,v_{p}) = \cC\{F_{1}(v_{1}),\dots,F_{p}(v_{p})\} =\cC(u_{1},\dots,u_{p}),         \label{eq:Copula A3} 
\ee
\vspace{-4ex}\\
where $u_{i} = F_{i}(v_{i}), i=1,\dots,p$.
If $\{V_i\}_{i=1}^{p}$ have marginal densities $\{f_{i}(v_i)\}_{i=1}^{p}$, 
then from (\ref{eq:Copula A3}) it follows that the joint density $f(v_{1},v_{2},\dots,v_{p}) $ is given by
\vspace{-4ex}\\
\be
f(v_{1},\dots,v_{p}) &= c(u_{1},\dots,u_{p})\prod_{i=1}^{p}f_{i}(v_i).  \label{eq:Copula A4}
\ee
\vspace{-4ex}\\
With $F_i(v_i) = u_{i} = H_i(x_{i}), i=1,\dots,p$, substitution of the copula density (\ref{eq:Copula A2}) into (\ref{eq:Copula A4}) gives 
\vspace{-6ex}\\
\be
&& f(v_{1},\dots,v_{p}) = c(u_{1},\dots,u_{p})\prod_{i=1}^{p}f_{i}(v_i) = \bigg\{\frac{h(x_{1},\dots,x_{p})}{\prod_{i=1}^{p}h_{i}(x_{i})}\bigg\}  \prod_{i=1}^{p}f_{i}(v_i). \label{eq:Copula A5}
\ee
\vspace{-4ex}\\

Equation (\ref{eq:Copula A3}) can be used to define flexible multivariate dependence structure using standard known multivariate densities \citeplatex{sklar1959}. 
Let $\MVN_{p}(\bmu,\bSigma)$ denote a $p$-variate normal distribution with mean vector $\bmu$ and positive semi-definite covariance matrix $\bSigma$. 
An important case is $\bX=(X_{1},\dots,X_{p})\trans \sim \MVN_{p}(\bzero,\bR)$, where $\bR$ is a correlation matrix. 
In this case, $\cC (u_{1},\dots,u_{p}|\bR) = \Phi_{p} \{\Phi^{-1}(u_{1}),\dots,\Phi^{-1}(u_{p})\vert \bR\}$, 
where $\Phi(x) = \Pr\{X \leq x \vert X\sim \Normal(0,1)\}$ and 
$\Phi_{p}(x_{1},\dots,x_{p}|\bR) = \Pr\{X_{1} \leq x_{1}, \dots,X_{p} \leq x_{p} \vert \bX \sim \MVN_{p}(\bzero,\bR)\}$. 
If $\bX \sim N_{p}(\bzero,\bSigma)$, where $\bSigma = ((\sigma_{i,j}))$ is a covariance matrix with $\sigma_{ii}=\sigma_i^2$, then defining $\bLambda=\diag(\sigma_{1}^2,\dots,\sigma_{p}^2)$ and $\bY = \bLambda^{-\frac{1}{2}}\bX$ and noting that $\bSigma=\bLambda^{1/2} \bR \bLambda^{1/2}$, we have  
\vspace{-4ex}\\
\bse
c(u_{1},\dots,u_{p}) = {\MVN_{p}(\bx\vert\bzero,\bSigma)}   /   {\MVN_{p}(\bx\vert\bzero,\bLambda)}
= |\bLambda|^{1/2} |\bSigma|^{-1/2} \exp\left\{-\bx\trans(\bSigma^{-1}-\bLambda^{-1})\bx/2\right\} \\
= |\bR|^{-1/2} \exp\{-\by\trans(\bR^{-1}-\bI_{p})\by/2\} 
=  {\MVN_{p}(\by\vert \bzero,\bR)}  /  {\MVN_{p}(\by\vert \bzero, \bI_{p})}.
\ese 
\vspace{-4ex}\\
Sticking to the standard normal case, 
a flexible dependence structure between random variables $\{V_i\}_{i=1}^{p}$ with given marginals $\{F_i(v_i)\}_{i=1}^{p}$ may thus be obtained  
assuming a Gaussian distribution on the latent random variables $\{Y_{i}\}_{i=1}^{p}$ 
obtained through the transformations $F_{i}(v_{i}) = u_{i} = \Phi(y_{i}), i=1,\dots,p$. 
The joint density of $\bV=(V_{1},\dots,V_{p})\trans$ is then given by
\vspace{-4ex}\\
\bse
&& \hspace{-1cm} f(v_{1},\dots,v_{p}) = c(u_{1},\dots,u_{p})\prod_{i=1}^{p}f_{i}(v_i) 
= \frac{\MVN_{p}(\by\vert \bzero,\bR)}   {\MVN_{p}(\by\vert \bzero, \bI_{p})}  \prod_{i=1}^{p}f_{i}(v_i). 
\ese
\vspace{-4ex}\\
We have 
\vspace{-6ex}\\
\bse
\Pr(V_{1} \leq v_{1},\dots,V_{p} \leq v_{p}) = \Pr[Y_{1} \leq \Phi^{-1}\{F_{1}(v_{1})\},\dots,Y_{p} \leq \Phi^{-1}\{F_{p}(v_{p})\}\vert \bY \sim \MVN_{p}(\bzero,\bR)].
\ese
\vspace{-4ex}\\
For $q\leq p$, with $(Y_{1},\dots,Y_{q})\trans \sim \MVN_{q}(\bzero,\bR_{q})$, we then have 
\vspace{-4ex}\\
\bse
\Pr(V_{1} \leq v_{1},\dots,V_{q} \leq v_{q}) = \Pr[Y_{1} \leq \Phi^{-1}\{F_{1}(v_{1})\},\dots,Y_{q} \leq \Phi^{-1}\{F_{q}(v_{q})\}\vert \bY \sim \MVN_{q}(\bzero,\bR_{q})],
\ese
\vspace{-4ex}\\
implying that the density of $(V_{1},\dots,V_{q})$ will be 
\vspace{-4ex}\\
\bse
&& \hspace{-1cm} f(v_{1},\dots,v_q) = c(u_{1},\dots,u_q)\prod_{i=1}^{q}f_{i}(v_i) 
= \frac{\MVN_{q}(\by\vert \bzero,\bR_{q})}   {\MVN_{q}(\by\vert \bzero, \bI_{q})}  \prod_{i=1}^{q}f_{i}(v_i). 
\ese
\vspace{-4ex}

\newpage 
\section{Hamiltonian Monte Carlo} \label{sec: HMC basics}
In this section, we present a brief review of the Hamiltonian Monte Carlo (HMC) sampling algorithm for easy reference.
Our presentation of HMC is inspired from \citelatex{neal2011mcmc} and \citelatex{betancourt2015hamiltonian}.
Let us assume that we want to draw samples of $q$ from $\pi(q)$. 
The HMC algorithm adds an additional momentum variable $p$ and considers following joint density
\vspace{-6ex}\\
\bse
\pi(p,q)=\pi(p\vert q)\pi(q).
\ese
\vspace{-4ex}\\
The Hamiltonian $\H(p,q)$ of a system is defined as the sum total of the kinetic energies and the potential energies of all the particles in the system.
Liouville's theorem on phase space equation from statistical mechanics states that the probability density of phase space is time-invariant along the trajectory. 
Based on that, it can be shown that for multivariate normal initial distribution the joint density 
\vspace{-4ex}\\
\bse
\pi(p,q)\propto \exp\{-\H(p,q)\}.
\ese 
\vspace{-4ex}\\
Thus we get 
\vspace{-4ex}\\
\bse
\H(p,q) = - \log\{\pi(p,q)\} 
=-\log\{\pi(p\vert q)\}-\log\{\pi(q)\}
=K(p\vert q)+U(q),
\ese
\vspace{-4ex}\\
where $K(p\vert q)$ is the kinetic energy and $U(q)$ is the potential energy. The first step of the HMC algorithm is to draw the momentum variable $p$ such that $p\sim\pi(p\vert q)$. The changes in $p$ and $q$ over time are governed by the following Hamiltonian's equations
\vspace{-4ex}\\
\bse
    & \frac{dq}{dt}=+\frac{\partial\H}{\partial p}=+\frac{\partial K}{\partial p},\\
    &\frac{dp}{dt}=-\frac{\partial\H}{\partial q}=-\frac{\partial K}{\partial q}-\frac{\partial U}{\partial q}.
\ese
\vspace{-4ex}\\
Computational implementation of these equations requires discretization of time, with small step size $\epsilon$ and evaluate the states at $t=\epsilon,2\epsilon,\ldots,$ so on.
\citelatex{neal2011mcmc} showed the superiority of the leapfrog method in producing excellent solution to a system of differential equations.
In most applications, including our own, $p$ is assumed to be independent of $q$ and is generated from a $\Normal(0,1)$ density.
The leapfrog method then proceeds as follows 
\vspace{-4ex}\\
\bse
    & p(t+\epsilon/2)=p(t) - (\epsilon/2)\frac{\partial U}{\partial q},\\
    & q(t+\epsilon)=q(t)+\epsilon(t+\epsilon/2),\\
    &p(t+\epsilon)=p(t+\epsilon/2)-(\epsilon/2)\frac{\partial U}{\partial q}.
\ese
\vspace{-4ex}\\
We iteratively compute the states at $t=\epsilon, 2\epsilon, 3\epsilon,\ldots,L\epsilon$, where $L$ stands for the number of leapfrog steps. 
\citelatex{neal2011mcmc} showed that if we take two choices of $L$ and $\epsilon$ such that $L_1\epsilon_1=L_2\epsilon_2$, the solution due to larger $L$ and smaller $\epsilon$ works better.
In our application, we keep $L$ fixed and tune $\epsilon$ within our MCMC implementation to achieve an acceptance rate between 0.6 to 0.9 which exhibits good mixing and efficiency in posterior sampling.
After running the iteration from $L$-many steps, we obtain an updated $q^{*}$ and $p^{*}$. 
The acceptance probability for the new $q^{*}$ is
\vspace{-4ex}\\
\bse
\exp\{U(q)-U(q^{*})+K(p)-K(p^{*})\},
\ese
\vspace{-4ex}\\
where $U(q)$ stands for the negative log-posterior and $K(p)=\frac{1}{2}p^{2}$ is the Gaussian kinetic energy term.

\newpage 
\section{Cubic B-splines}\label{sec: cubic B-splines}

Consider knot-points $t_{-2} = t_{-1} = t_{0} = t_{1}= A < t_{2} < \dots < B = t_{K} = t_{K+1} = t_{K+2}=t_{K+3}$,
where $t_{1:K}$ are equidistant with $\delta = (t_{2} - t_{1})$.
For $j=1,2,\dots,(K+3)$, cubic B-splines $b_{3,j}$ are defined as 
\vspace{-4ex}\\
\bse
 b_{3,j}(x) &= \left\{\begin{array}{ll}
        \frac{1}{6\delta^{3}}\{(x-t_{j-2})\}^{3}  				& ~~~~\text{if } t_{j-2} \leq x < t_{j-1},  \\
        \frac{1}{6\delta^{3}}\{\delta^{3}+3(x-t_{j-1})\delta^2+3(x-t_{j-1})^2\delta-3(x-t_{j-1})^3\}	     	& ~~~~\text{if } t_{j-1} \leq x < t_{j},  \\
        \frac{1}{6\delta^{3}}\{\delta^{3}+3(t_{j+1}-x)\delta^2+3(t_{j+1}-x)^2\delta-3(t_{j+1}-x)^3\}		       	    	& ~~~~\text{if } t_{j} \leq x < t_{j+1},  \\
        \frac{1}{6\delta^{3}}\{(t_{j+2}-x)\}^{3}  		       	    	& ~~~~\text{if } t_{j+1} \leq x < t_{j+2},  \\
        0  							& ~~~~ \text{otherwise}.
        \end{array}\right.
\ese

\vspace{-3ex}
\begin{figure}[!h]
\begin{center}
\includegraphics[width = 1\textwidth, trim=1cm 1.25cm 1cm 0.5cm , clip=true]{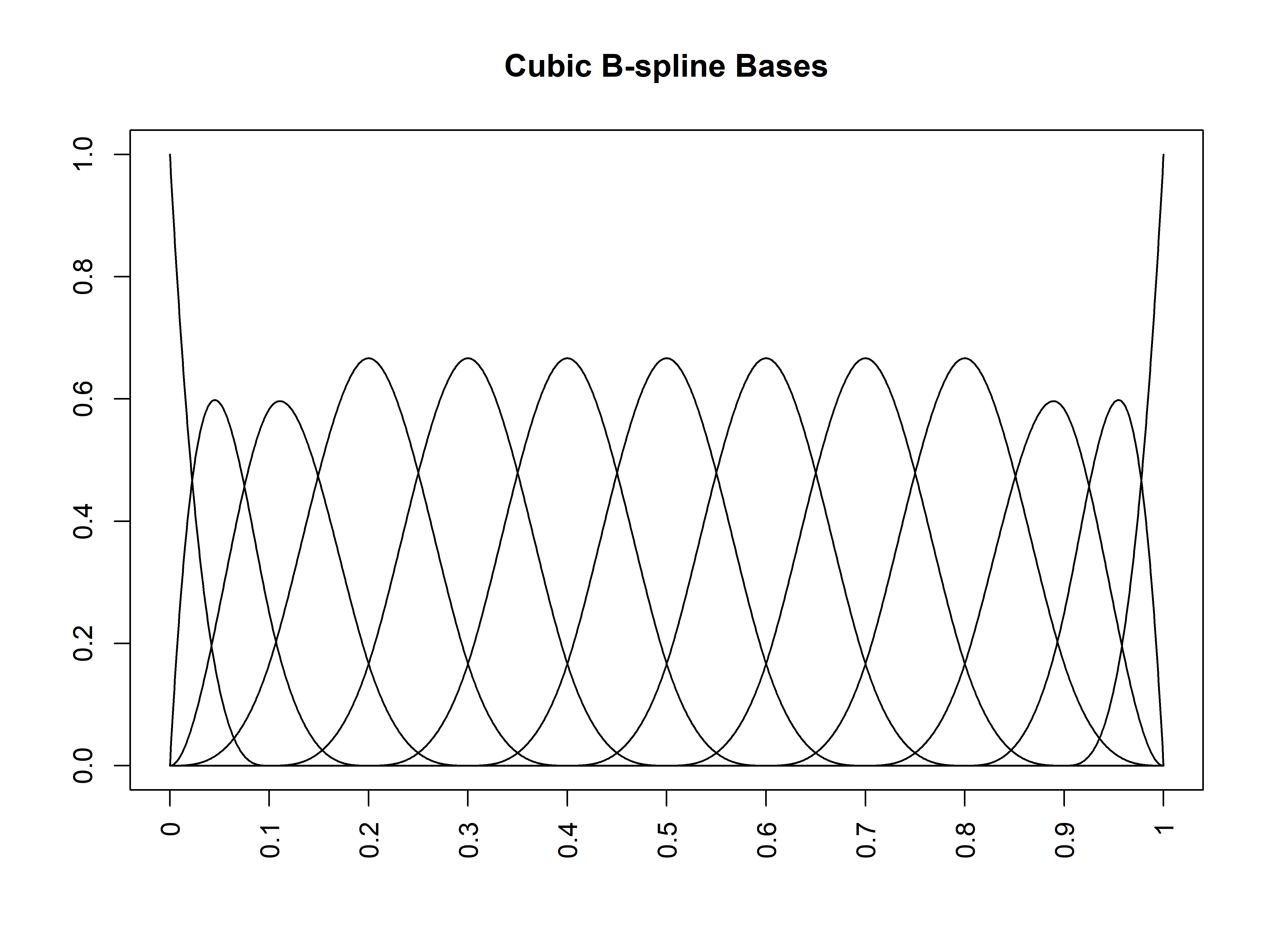}
\end{center}
\caption{\baselineskip=10pt Plot of 13 cubic B-splines on $[0,1]$ defined using $10$ knot points that divide $[0,1]$ into $K=10$ equal sub-intervals.}
\label{fig: cubic B-splines}
\end{figure}

\newpage
\section{Model Identification} \label{sec: identifiability}
The additive zero mean error formulation of our model in Section 3.3 of the main paper 
readily establishes the identifiability of $f_{\bx}$ when the joint and conditional distributions of $\bw,\bx$ are all bounded, the characteristic function of $(\bx\vert\bw)$ is non-vanishing everywhere and $m_{i}\geq 3$ replicates are available for each subject $i$ \citeplatex{hu2008instrumental}. 
Broadly speaking, if the density $f_{\bx\vert\bw}$ varies with $\bx$, its characteristic function does not vanish. 
With sufficient variability of the density of $\bx\vert\bw$, the observations $\bw$ have enough information to allow the recovery of the density of $\bx$ \citeplatex{sarkar2018bayesian}.

\section{Posterior Computation} \label{sec: detailpost comp}

We now discuss our MCMC algorithm to draw posterior samples of the model parameters. 
We shall provide full conditionals for all the parameters.
In addition, we also provide explicit expressions of the derivatives for the parameters that are updated using HMC.

\vspace*{-10pt}
\subsubsection*{\bf Choice of Hyper-parameters and MCMC Initial Values} 

The starting values of some of the parameters for the multivariate problem are determined by first fitting the univariate sub-model of \citelatex{sarkar2021bayesian}.
We describe the hyper-parameter choices and the initial values for the sampler for the marginal univariate models first. 
Unless otherwise explicitly specified, the prior hyper-parameter choices for similar model components for the multivariate model remain the same as that used for the univariate models. 
We only detail the sampling steps for the multivariate method. 
To make the recalls for all the components to be unit free and have shared support, we transformed the recalls as 
$w_{\ell,i,j} = 20 \times \frac{w_{\ell,i,j}}{\max\{w_{\ell,i,j}\}}$.
The latent $x_{\ell,i}$'s can then be safely assumed to lie in $[0,10]$, greatly simplifying model specification and hyper-parameter selection. 

For the univariate samplers, we then used the subject-specific sample means $\overline{w}_{\ell,1:n}$ as the starting values for $x_{\ell,1:n}$. 
The appropriate number of mixture components in a mixture model depends on the flexibility of the component mixture kernels as well as on the specific demands of the particular application at hand. 
With appropriately chosen mixture kernels, univariate mixture models with 5-10 components have often been found to be sufficiently flexible. 
Similar observations have also been made for penalized mixtures of splines \citeplatex{ruppert2002selecting}. 
Detailed guidelines on selecting the number of mixture components for the specific context of deconvolution problems can be found in Sections S.1 and S.6 in the supplementary material of \citelatex{sarkar2018bayesian}. 
Based on these guidelines and extensive numerical experiments, 
we used $10$ equidistant knot points for the B-splines supported on $[A,B] = [0,10]$ for modeling the functions $s^{2}(\norm{\bx}/d)$ and $\kappa_{\ell}(x_{\ell})$. 
Thus there are $K_{s}=K_{\kappa}=13$ many basis functions. 
We used $K=10$ mixture components for the truncated normal mixtures modeling their densities. 
For the Dirichlet prior hyper-parameter, we set $\alpha = 1/K$. 
The hyper-parameters for the smoothness-inducing parameters are set to be mildly informative as $a_{\xi}=a_{\beta}=a_{\vartheta}=10, b_{\vartheta}=b_{\beta}=b_{\xi}=1$. 
Introducing latent mixture component allocation variables $\bc_{x,1:d,1:n}$, $\bc_{\epsilon,1:d,1:N}$ and $\bc2_{\epsilon,1:d,1:N}$,
we can write the univariate sub-model of \citelatex{sarkar2021bayesian} as 
\vspace{-5ex}\\
\bse
& (x_{\ell,i} \vert c_{x,\ell,i}=k,\mu_{x,\ell,k},\sigma_{x,\ell,k}^{2}) \sim \TN(x_{\ell,i} \vert \mu_{x,\ell,k},\sigma_{x,\ell,k}^{2},[A,B]), ~~~\ell=1,\dots,d,~~\hbox{and}\\
& (\epsilon_{\ell,i,j} \vert c_{\epsilon,\ell,i,j}=k,c2_{\epsilon,\ell,i,j}=t,\mu_{\epsilon,\ell,k,t},\sigma_{\epsilon,\ell,k,t}^{2}) \sim \Normal(\epsilon_{\ell,i,j} \vert \mu_{\epsilon,\ell,k,t},\sigma_{\epsilon,\ell,k,t}^{2}),\\
&\hspace{11cm}~\ell=1,\dots,d.
\ese
\vspace{-5ex}

The mixture labels $c_{x,\ell,i}$'s, and the component-specific parameters $\mu_{x,\ell,k}$'s and $\sigma_{x,\ell,k}$'s are initialized by fitting a $k$-means algorithm with $k=K$.
The parameters of the distribution of scaled errors are initialized at values that correspond to the special standard normal case.
The initial values of the smoothness-inducing parameters are set at $\sigma_{\vartheta,\ell}^{2} = \sigma_{\xi,\ell}^{2} = \sigma_{\xi,\ell}^{2} = 0.1$. 
The associated mixture labels $c_{\epsilon,\ell,i,j}$'s are thus all initialized at $c_{\epsilon,\ell,i,j}=1$.
The initial values of $\bvartheta_{\ell}$'s are obtained by maximizing 
\vspace{-5ex}\\
\bse
\ell(\bvartheta_{\ell}\vert \sigma_{\vartheta,\ell}^{2},\overline{\bw}_{\ell,1:n}) = -\frac{\bvartheta_{\ell}\trans \bP_{\ell} \bvartheta_{\ell}}{2\sigma_{\vartheta,\ell}^{2}} - \sum_{i=1}^{n} \frac{1}{2 s_{\ell}^{2}(\overline{w}_{\ell,i},\bvartheta_{\ell})} \sum_{j=1}^{m_{i}}(w_{\ell,i,j}-\overline{w}_{\ell,i})^{2}
\ese
\vspace{-5ex}\\
with respect to $\bvartheta_{\ell}$. 

We now discuss how we set the initial values of the sampler for the multivariate method. 
The starting values of the $x_{\ell,i}$'s were all set at the corresponding estimates returned by the univariate samplers. 
The initialization of the B-spline coefficients of $\kappa_{\ell}(\cdot)$'s and $s^{2}(\cdot)$ were done based on the estimated $x_{\ell,i}$'s from the univariate sampler as described below. 
We first normalize $\bw_{i,j}$ as $\wt{\bw}_{i,j}=\bw_{i,j}/\|\bw_{i,j}\|_{2}$. 
Similarly, we also obtained $\wt{\bx}_{i}$'s. 
Then, for each individual $i$, 
we assume $\wt{\bw}_{i,j}\sim\vMF(\bFf_{i})$, where vMF stands for the von 
Mises-Fisher distribution \citeplatex{jupp1979maximum} 
and obtain maximum likelihood estimates $\wh{\bFf}_{i} = (\wh{f}_{1,i},\dots,\wh{f}_{d,i})\trans$ using the R package {\tt movMF} \citeplatex{hornik2014movmf}. 
Based on these estimates we set $\kappa_{\ell,i}(x_{\ell,i})=\wh{f}_{\ell,i}/\wt{x}_{\ell,i}$. 
Similarly for each individual, we calculate the variance of $v_{i}=\left\{\log\left(\frac{\|\bw_{i,j}\|_{2}}{\|\bx_{i}\|_{2}}\right), j=1,\ldots,m_{i}\right\}$ and set $s^{2}(\|\bx_{i}\|_{2}/d)=v_{i}$.
Using the R package {\tt nnls} \citeplatex{mullen2007nnls}, we then fit the non-negative least squares to obtain the B-spline coefficients $\bbeta_{\kappa,\ell}$ and $\bbeta_s$ for cubic splines with 10 knots.
For $\bbeta_{\kappa,\ell}$, the first input in {\tt nnls} is the matrix $\bM_{\ell}$ of dimension $n\times 13$, where $M_{\ell,i,i'}=B_{i'}(x_{\ell,i})$ and the second input is $\{\kappa_{\ell,i'}(x_{\ell,i}): i=1,\ldots,n\}$. Similarly, to estimate $\bbeta_{s}$, the corresponding first matrix input $\bM_s$ is of dimension $n\times 13$, where $M_{s,i,i'}=B_{i'}(\|\bx_{i}\|_{2}/d)$ and the vector input is $\{s^{2}(\|\bx_{i}\|_{2}/d): i=1,\ldots,n\}$.

We set the number of shared atoms of the mixture models for the densities $f_{x,\ell}$ at $K = 10$. 
The shared atoms of the mixtures of truncated normals for the marginal densities $f_{\bx,\ell}$ are initialized by iteratively sampling them from their posterior full conditionals $500$ times, 
keeping the $x_{\ell,i}$'s fixed at their estimated initial values.   

Finally, the parameters specifying $\bR_{\bx}$ were set at values that correspond to the special case $\bR_{\bx}=\bI_{p}$. 
In our sampler for the multivariate problem, 
we first update the parameters specifying the different marginal densities using a pseudo-likelihood that ignores the contribution of the copula. 
The parameters characterizing the copula and the latent $\bx_{i}$'s are then updated using the exact likelihood function conditionally on the parameters obtained in the first step. 
The results of \citelatex{dos2008copula} suggest that such two-stage approach performs just as good as joint estimation procedures, validating their use for computational simplicity.
We then update the parameters of the marginal densities again and so forth. 
Lastly, we update the B-spline coefficients involved in our measurement error model using an HMC sampler.

\vspace*{-10pt}

\subsubsection*{\bf MCMC Iterations}

Our sampler for the multivariate model iterates between the following steps. 

\begin{enumerate}[topsep=0ex,itemsep=2ex,partopsep=2ex,parsep=0ex, leftmargin=0cm, rightmargin=0cm, wide=3ex]

\item {\bf Updating the parameters specifying $f_{x,\ell}, \ell=1,\dots,d$:} We modelled the marginal densities of the components $\ell=1,\dots,d$ using mixtures of truncated normals with shared atoms. We update the mean and variance of the mixture normal components using HMC. The mixing probabilities and mixture indicators are updated using the full conditional conjugate posterior distributions.

Specifically, the full conditional of $\pi_{x,\ell,k}$ is given by
\vspace{-5ex}\\
\bse
p(\bpi_{x,\ell} \vert \bzeta) &=& \textstyle \Dir\{\alpha_{x,\ell}+n_{x,\ell}(1),\dots,\alpha_{x,\ell}+n_{x,\ell}(K)\}.
\ese
\vspace{-5ex}\\
where $n_{x,\ell}(k) = \sum_{i=1}^{n}1(c_{x,\ell,i}=k)$ as before.
The full conditional of $c_{x,\ell,i}$ is given by  
\vspace{-5ex}\\
\bse
p(c_{x,\ell,i}=k \vert \bzeta) &\propto& \pi_{x,\ell,k} \times \TN(x_{\ell,i}\vert \mu_{x,k},\sigma_{x,k}^{2},[A,B]), 
\ese
\vspace{-5ex}\\
a standard multinomial. 
The full conditional of  $\mu_{x,k}$ is given by
\vspace{-5ex}\\
\bse
\textstyle p(\mu_{x,k}\vert \bzeta) \propto p_{0}(\mu_{x,k}) \times \prod_{\ell=1}^{d}\prod_{\{i: c_{x,\ell,i}=k\}}\TN(x_{\ell,i}\vert \mu_{x,k},\sigma_{x,k}^{2},[A,B]),
\ese
\vspace{-5ex}\\
which gives the following negative log-likelihood
\vspace{-5ex}\\
\bse
&\frac{1}{2}\frac{(\mu_{x,k}-\mu_{X0})^{2}}{\sigma_{x0}^{2}} + \frac{1}{2}\sum_{\ell=1}^{d}\sum_{\{i: c_{x,\ell,i}=k\}}\frac{(x_{\ell,i}- \mu_{x,k})^{2}}{\sigma_{x,k}^{2}} \\&\quad+{|\{i: c_{x,\ell,i}=k\}|}\log\left\{\Phi\left(\frac{B- \mu_{x,k}}{\sigma_{x,k}}\right)-\Phi\left(\frac{A- \mu_{x,k}}{\sigma_{x,k}}\right)\right\},
\ese
\vspace{-5ex}\\
where $\Phi(\cdot)$ stands for the standard normal cdf and the derivative is given by
\vspace{-5ex}\\
\bse
&\frac{(\mu_{x,k}-\mu_{x0})}{\sigma_{x0}^{2}} - \sum_{\ell=1}^{d}\sum_{\{i: c_{x,\ell,i}=k\}}\frac{(x_{\ell,i}- \mu_{x,k})}{\sigma_{x,k}^{2}} - \frac{|\{i: c_{x,\ell,i}=k\}|\left\{\phi\left(\frac{B- \mu_{x,k}}{\sigma_{x,k}}\right)-\phi\left(\frac{A- \mu_{x,k}}{\sigma_{x,k}}\right)\right\}}{\sigma_{x,k}\left\{\Phi\left(\frac{B- \mu_{x,k}}{\sigma_{x,k}}\right)-\Phi\left(\frac{A- \mu_{x,k}}{\sigma_{x,k}}\right)\right\}}.
\ese
\vspace{-5ex}\\
Similarly, the full conditional of  $\sigma_{x,k}^{2}$ is given by
\vspace{-5ex}\\
\bse
\textstyle p(\sigma_{x,k}^{2} \vert \bzeta) \propto p_{0}(\sigma_{x,k}^{2}) \times \prod_{\ell=1}^{d} \prod_{\{i: c_{x,\ell,i}=k\}}\TN(x_{\ell,i}\vert \mu_{x,k},\sigma_{x,k}^{2},[A,B]).
\ese
\vspace{-5ex}\\
which gives the following negative log-likelihood
\vspace{-5ex}\\
\bse
&2\log(\sigma_{x,k}^{2}) + \sigma_{x,k}^{-2} + \frac{1}{2}\sum_{\ell=1}^{d}\sum_{\{i: c_{x,\ell,i}=k\}}\frac{(x_{\ell,i}- \mu_{x,k})^{2}}{\sigma_{x,k}^{2}} \\&\quad + {|\{i: c_{x,\ell,i}=k\}|}\log\left\{\Phi\left(\frac{B- \mu_{x,k}}{\sigma_{x,k}}\right)-\Phi\left(\frac{A- \mu_{x,k}}{\sigma_{x,k}}\right)\right\},
\ese
\vspace{-5ex}\\
and the derivative with respect to $\sigma_{x,k}^{2}$ is given by
\vspace{-5ex}\\
\bse
&2/\sigma_{x,k}^{2}- \sigma_{x,k}^{-4} - \frac{1}{2}\sum_{\ell=1}^{d}\sum_{\{i: c_{x,\ell,i}=k\}}\frac{(x_{\ell,i}- \mu_{x,k})^{2}}{\sigma_{x,k}^{4}}+|\{i: c_{x,\ell,i}=k\}|/(2\sigma_{x,k}^2)\\&\quad- \frac{|\{i: c_{x,\ell,i}=k\}|\left\{\phi\left(\frac{B- \mu_{x,k}}{\sigma_{x,k}}\right)(B- \mu_{x,k})-\phi\left(\frac{A- \mu_{x,k}}{\sigma_{x,k}}\right)(A- \mu_{x,k})\right\}}{2\sigma_{x,k}^3\left\{\Phi\left(\frac{B- \mu_{x,k}}{\sigma_{x,k}}\right)-\Phi\left(\frac{A- \mu_{x,k}}{\sigma_{x,k}}\right)\right\}}.
\ese
\vspace{-3ex}\\
These parameters are updated by HMC with Gaussian kinetic energy as described in Section~\ref{sec: HMC basics} in the supplementary materials.

\item {\bf Updating the B-spline coefficients specifying $s^{2}$ and $\kappa_{\ell}$ for $\ell=1,\dots,d$: } 
These parameters are also updated using HMC sampler. 
We provide below the associated negative log-likelihood functions and the corresponding derivatives.
\begin{enumerate}
    \item 
The negative log-likelihood to update the B-spline coefficients $\bbeta_{s}$ is
\vspace{-5ex}\\
\bse
L_{neg,1} = \sum_{i=1}^{n} \frac{m_{i}}{2} \log\{s^{2}(\|\bx_{i}\|_{2}/d)\} + \frac{1}{2}\sum_{i=1}^{n}\sum_{j=1}^{m_{i}}\frac{\{\log(\|\bC_{i}^{-1}\bw_{i,j}\|_{2}/\|\bx_{i}\|_{2})+s^{2}(\|\bx_{i}\|_{2}/d)/2\}^{2}}{s^{2}(\|\bx_{i}\|_{2}/d)} \\ 
+ \frac{1}{2} \sum_{k=1}^{K_{s}} \frac{(\beta_{s,k}-\mu_{s,k})^{2}}{\sigma_{s}^{2}} 1_{\beta_{s,k}\in[0,\infty]}.
\ese
\vspace{-3ex}\\
And the derivative is 
\vspace{-5ex}\\
\bse
&\frac{\partial L_{neg,1}}{\partial \beta_{s, k}}=\sum_{i=1}^{n}B_{k}(\|\bx_{i}\|_{2}/d) \sum_{j=1}^{m_i}\bigg[\frac{1}{2 s^{2}(\|\bx_{i}\|_{2}/d)} + \frac{\log(\|\bC_{i}^{-1}\bw_{i,j}\|_{2}/\|\bx_{i}\|_{2})+s^{2}(\|\bx_{i}\|_{2}/d)/2}{2 s^{2}(\|\bx_{i}\|_{2}/d)} \\
&\qquad - \frac{\big\{\log(\|\bC_{i}^{-1}\bw_{i,j}\|_{2}/\|\bx_{i}\|_{2})+s^{2}(\|\bx_{i}\|_{2}/d)/2 \big\}^2}{2 s^{4}(\|\bx_{i}\|_{2}/d)}\bigg] +  \frac{(\beta_{s,k}-\mu_{s,k})}{\sigma_{s}^{2}}{1}_{\beta_{s,k}\in[0,\infty]}.
\ese
\vspace{-5ex}

\item The negative log-likelihood to update the B-spline coefficients $\bbeta_{\kappa,\ell}$ is
\vspace{-5ex}\\
\bse
L_{neg,2}=\sum_{i=1}^{n}\sum_{j=1}^{m_{i}}\left[\log\{M(\bF_{i})\}-\trace\left\{\mathcal{G}\left(\bC_{i}^{-1}\bw_{i,j},\bx_{i}\right) \bF_{i}\right\}\right] + \sum_{i=1}^{n}\sum_{j=1}^{m_{i}}\log(\|\bC_{i}^{-1}\bw_{i,j}\|_{2})\\
+ \frac{1}{2}\sum_{i=1}^{n}\sum_{j=1}^{m_i}\frac{\{\log(\|\bC_{i}^{-1}\bw_{i,j}\|_{2}/\|\bx_{i}\|_{2})+s^{2}(\|\bx_{i}\|_{2}/d)/2\}^{2}}{s^{2}(\|\bx_{i}\|_{2}/d)} 
+ \frac{1}{2}\sum_{k=1}^{K_{\kappa}} \frac{(\beta_{\kappa,\ell,k}-\mu_{\kappa,\ell,k})^{2}}{\sigma_{\kappa,\ell}^{2}}{1}_{\beta_{\kappa,\ell,k}\in[0,\infty]}.
\ese
\vspace{-5ex}\\
We consider adaptive random-walk MH sampling for $\bbeta_{\kappa,\ell}$ using a multivariate normal proposal with a covariance matrix being set based on the generated posterior samples adaptively.

\end{enumerate}

\item {\bf Updating the values of $\bx$:}
The full conditionals for $\bx_{i}$ are given by
\vspace{-5ex}\\
\bse
&&\hspace{-1cm}(\bx_{i}\vert \bzeta) 
\propto f_{\bx}(\bx_{i}\vert \bzeta) \times \textstyle\prod_{j=1}^{m_{i}} f_{\bw \vert {\bx}}(\bw_{i,j} \vert \bx_{i}, \bzeta)  \\
&& \textstyle \propto |\bR_{\bx}|^{-1/2} \exp\left\{-\frac{1}{2}\by_{\bx,i}\trans(\bR_{\bx}^{-1}-\bI_{d})\by_{\bx,i}\right\}\prod_{\ell=1}^{d}f_{x,\ell}(x_{\ell,i}\vert \bzeta) \times \\
&&\hspace*{-2cm} \prod_{j}\frac{1}{M(\bF_{i})}\etr\left\{\mathcal{G}\left(\bC_{i}^{-1}\bw_{i,j},\bx_{i}\right) \bF_{i}\right\} \frac{\|\bx_{i}\|_{2}}{\|\bC_{i}^{-1}\bw_{i,j}\|_{2} s(\|\bx_{i}\|_{2}/d)} \exp\left[-\frac{\left\{\log(\|\bC_{i}^{-1}\bw_{i,j}\|_{2}/\|\bx_{i}\|_{2}) + s^{2}(\|\bx_{i}\|_{2}/d)/2\right\}^{2}}{2s^{2}(\|\bx_{i}\|_{2}/d)}\right],
\ese
\vspace{-3ex}\\
where $F_{x,\ell}(x_{\ell,i}\vert\bzeta)=\Phi(y_{x,\ell,i})$. 
These full conditionals do not have closed forms. 
Metropolis-Hastings (M-H) steps with independent truncated normal proposals for each component are used within the Gibbs sampler.

\item {\bf Updating the parameters specifying the copula:} 
We have $F_{x,\ell}(x_{\ell,i}\vert\bzeta)=\Phi(y_{x,\ell,i})$ for all $i=1,\dots,n$ and $\ell=1,\dots,d$.
Conditionally on the parameters  specifying the marginals, $\by_{\bx,1:d,1:n}$ are thus known quantities. 
We plug-in these values and use that $(\by_{\bx,i}\vert\bR_{\bx}) \sim \MVN_{d}(\bzero,\bR_{\bx})$ to update $\bR_{\bx}$.  
The full conditionals of the parameters specifying $\bR_{\bx}$ do not have closed forms. 
We use random walk M-H steps to update the $d(d-1)/2$ set of polar angles. 
Since polar angles have bounded support, they are proposals generated from truncated normals, supported according to the respective polar angles. 
The angles $\{\zeta_{m,m-1}:m=1,\ldots,d\}$ are supported in $[0, 2\pi]$, whereas other angles are supported in $[0,\pi]$. 
We discretized the set of possible values of each $\zeta_{m,s}$ with $M=41$ equidistant grid points covering its support. 
A new value $\zeta_{s,new}$ is proposed at random from the set comprising the current value and its two neighbors. 
Their proposed values are accepted with probabilities $\min\{1,a(\zeta_{m,s,new})/a(\zeta_{m,s})\}$, where 
\vspace{-4ex}\\
\bse
\textstyle a(\zeta_{m,s}) = \det(\bSigma_{\bx}(\zeta_{m,s}))^{-d/2}\exp\left\{- (1/2)\sum_{i=1}^{n}\sum_{j=1}^{m_{i}}\by_{\bx,i,j}\trans\{\bSigma_{\bx}(\zeta_{x,s})\}^{-1}\by_{\bx,i,j}\right\}. 
\ese
\vspace{-7ex}

\end{enumerate}

With carefully chosen initial values and proposal densities for the M-H steps, we were able to achieve quick convergence for the MCMC samplers. 
We also automatically tune the step lengths for all the Metropolis proposals to achieve acceptance rates within a pre-specified range, namely, 0.4-0.5 for the M-H steps and 0.6-0.9 for HMC.
For our proposed method, $5,000$ MCMC iterations were run in each case with the initial $3,000$ iterations discarded as burn-in. 
The remaining samples were further thinned by a thinning interval of $5$. 
We programmed in {R}.
With $n=1000$ subjects and $m_{i}=3$ proxies for each subject, on an ordinary laptop, $5,000$ MCMC iterations required approximately $4$ hours to run.

\section{Computational Cost and Complexity}
\begin{figure}[!hb]
    \centering
    \includegraphics[width = 0.75\textwidth]{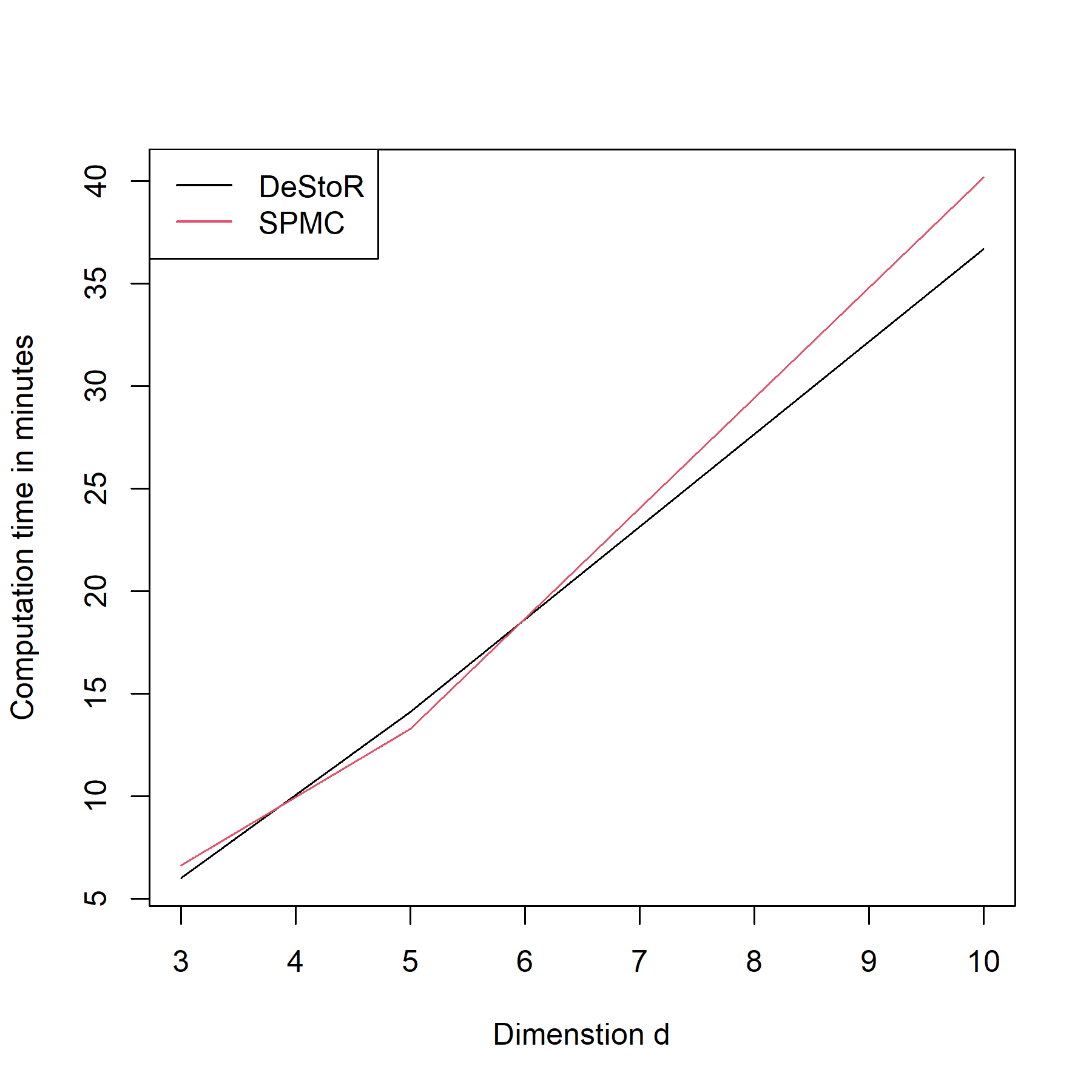}
    \caption{Computation time with increasing dimension $d$.}
    \label{fig:comput-time}
\end{figure}

We also determine the time complexity of the proposed method in the increasing $d$ regime via simulations. 
Specifically, we run the proposed method for three choices of $d$, namely $d=3, 5, 10$, and $30$ simulation replications. 
We repeat the same steps for SPMC as well.
For a fair comparison, we run both of the two methods for 2000 MCMC iterations.
Our final computation times are averaged over all $30$ replications. 
Subsequently, we fit the model $\log(t)$ on $\log(d)$ to compute $a$ such that $t=O(d^a)$, where $t$ is the computation time when the data is $d$-dimensional. 
The results are shown in Figure~\ref{fig:comput-time}.
The solution of $a$ turns out to be 1.49 for our method. 
For SPMC, the order of computation turns out to be 1.51 which is almost the same as our proposed method. 
However, our proposed method converges faster and requires fewer MCMC steps, likely due to the use of more efficient gradient-based MH proposals for some of the parameters.

\clearpage\newpage
\section{Posterior Consistency} \label{sec: asymptotics}
We study the posterior consistency of the proposed model as the sample size increases to infinity. We make following assumption for true distribution of $f_{0,\bx}$ 
\bse
    \textstyle f_{0,\bx}(\bx)=\bR_{0,\bx}^{-1}\exp\left\{-\by_{\bx}\trans(\bR_{0,\bx}^{-1}-\bI_{d})\by_{\bx}\right\}\prod_{\ell=1}^{d} f_{0,x,\ell}(x_{\ell}),
\ese
with $F_{0,x,{\ell}}(x_{\ell})=\Phi(y_{x,{\ell}})$, where $F_{0,x,{\ell}}$ is the cdf corresponding to $f_{0,x,{\ell}}$, and the univariate densities $f_{0,x,{\ell}}(\cdot)$'s are supported in $[A,B]$ such that $f_{0,x,{\ell}}(\cdot)\in\F$ where $\mathcal{F}$ is the class of square-integrable functions.
Convex combinations of Gaussians are dense in $L_{2}$ meaning they can approximate any square-integrable function with arbitrarily small approximation error \citep{kostantinos2000gaussian}.
One may vary the support of $f_{0,\bx,\ell}$ with $\ell$, for example, the support for $x_{\ell}$ may be $[A_{\ell},B_{\ell}]$. 
However, choosing $A=\min_{\ell} A_{\ell}$ and $B=\max_{\ell} B_{\ell}$,
the proofs will remain the same with minor technical modifications.
We present here the main results. 
To save space, detailed proofs are deferred to Section \ref{sec: proofs} of the supplementary materials. 
Let $\btheta = \{\kappa_{\ell}(\cdot):\ell=1,\ldots,d,s^{2}(\cdot),K_{\kappa},K_s\}\bigcup \{K,\bR_{\bx},f_{\bx,\ell}:\ell=1,\ldots,d\}=(\btheta',\btheta'')$ 
denote the complete set of parameters. 
The marginal density for $\bw$ is $f_{\bw}(\bw)=\int_{\X}f_{\bw\vert\bx}(\bw\vert\bx)f_{\bx}(\bx)d\bx$. 
We have $\bx \in \X = [A,B]^{d}$. 
Also, let the replicates $\bw\in\mathcal{W}\subseteq\mathbb{R}^d$.
We have $F_{0,\bx}(\X)=1$ and $F_{\bx}(\X)=1$. 
Let $\btheta_{0}=(\btheta'_{0},\btheta''_{0})$ be the true value of $\btheta$.
We make the following set of assumptions. 

\begin{Assmp} \label{assmp: KL w}
 $\int f_{0,\bw}(\bw)\log\frac{\int f_{0,\bw\vert\bx}(\bw\vert\bx)f_{0,\bx}(\bx)d\bx}{\inf_{\bx}f_{0,\bw\vert\bx}(\bw\vert\bx)}d\bw < \infty$.
\end{Assmp}
\begin{Assmp}\label{assmp: lower}
 For any compact set $\bw_{0}\subset\mathcal{W}$, $e:=\inf_{\bw\in \bW_{0}}\int f_{0,\bw\vert\bx}(\bw\vert\bx)f_{0,\bx}(\bx)d\bx>0$.
\end{Assmp}

\begin{Assmp}\label{assmp: func}
 The coefficient functions $\kappa_{0,\ell}(\cdot)$'s are H\"older smooth with regularity $\iota$ and the function $s_{0}^{2}(\cdot)$ is also H\"older smooth with regularity $\iota'$. We further assume $0<m_{\kappa}<\kappa_{0,\ell}(\cdot)<M_{\kappa}<\infty$ for all $\ell$ and $0<m_{s}<s_{0}^{2}(\cdot)<M_{s}<\infty$.
\end{Assmp}

Assumption~\ref{assmp: KL w} and Assumption~\ref{assmp: lower} are in line with other existing posterior consistency results for mixture models \citeplatex{ghosal1999posterior, wu2008kullback}. 
Assumption~\ref{assmp: func} ensures that for some $0<M<\infty$, $\sup\{f_{0,\bw\vert\bx}(\bw\vert\bx): (\bw,\bx) \in \W_{0} \times \X\}<eM$.

For the sake of generality, 
we also put priors on $K_{\kappa}$, $K_s$ and $K$ with probability mass functions given by 
\bse\label{eq:bij}
&\Pi(K_{\kappa}=k)=b_{\kappa 1}\exp\{-b_{\kappa 2} k (\log k)^{b_{\kappa 3}}\},
~~~\Pi(K_{s}=k)=b_{s 1}\exp\{-b_{s 2} k (\log k)^{b_{s 3}}\},\\
&\text{and}~\Pi(K_{}=k)=b_{1}\exp\{-b_{2} k (\log k)^{b_{3}}\},
\ese
where $b_{\kappa1},b_{\kappa2},b_{s1},b_{s2},b_{1},b_{2}>0$ and $0\le b_{\kappa3},b_{s3},b_{3}\le 1$ for $i=1,2$. 
As special cases of the above, we can obtain the Poisson and geometric probability mass functions respectively for $b_{\kappa3},b_{s3},b_{3}=1$ or $0$. 
However, for computational simplicity, we do not consider these priors while fitting the model but propose to tune $K_{\kappa}, K_{s}$ and $K$ based on the sample size. 

Posterior consistency theory studies recovery of the `true' parameter $\btheta_{0}$ with increasing sample size when the data is sampled from the distribution characterized by $\btheta_{0}$. Our notion of recovery is based on following $L_{1}$-distance metric 
\bse
d(\btheta_{1},\btheta_{2}) = \|f_{1,\bw}-f_{2,\bw}\|=\int|f_{1,\bw}-f_{2,\bw}|d\bw.
\ese

\begin{Thm}
Under Assumptions 1-4, $\Pi_{n}\{\theta:d(\btheta,\btheta_{0})>\epsilon \vert \bw_{1:n}\} \rightarrow 0$ as $n\rightarrow\infty$ almost surely in $P_{f_{0,\bw}}$ for every $\epsilon>0$.
\end{Thm}

The proof is based on Theorem 2 of \citelatex{ghosal1999posterior}. 
In the context of density deconvolution, a slightly more appealing notion of recovery could be in terms of a distance metric involving $\btheta_{0}''$ alone. 
To show posterior consistency in terms of a distance metric involving $\btheta_{0}''$ alone such as the Wasserstein metric as in \citelatex{gao2016posterior,su2020nonparametric}, we need to establish an inversion inequality.
However, such inversion inequality is difficult to formulate unless $\btheta'$ is known.
Since, our proposed model does not assume $\btheta'$ to be known, establishing such consistency result in our setting is beyond the scope of this paper. 
Our notion of recovery thus involves the complete set of parameters that focuses on the estimation of all parameters simultaneously.

\newpage
\section{Proofs of Theoretical Results} \label{sec: proofs}

We first establish that, under the assumptions described in Section \ref{sec: asymptotics}, the truth belongs to the Kullback-Leibler (KL) support of the prior. 
We rewrite the KL as
\vspace{-5ex}\\
\bse
    &&\hskip -35pt \textrm{KL}\{f_{0,\bw}(\bw),f_{\bw}(\bw)\}=\large\int\log\left\{\frac{\int_{\X}f_{0,\bw\vert\bx}(\bw\vert\bx)f_{0,\bx}(\bx)d\bx}{\int_{\X}f_{\bw\vert\bx}(\bw\vert\bx)f_{\bx}(\bx)d\bx}\right\}f_{0,\bw}(\bw)d\bw\nonumber\\
    &&\hskip -35pt = \footnotesize{{\int\log\left\{\frac{\int_{\X}f_{0,\bw\vert\bx}(\bw\vert\bx)f_{0,\bx}(\bx)d\bx}{\int_{\X}f_{0,\bw\vert\bx}(\bw\vert\bx)f_{\bx}(\bx)d\bx}\right\}f_{0,\bw}(\bw)d\bw+\int\log\left\{\frac{\int_{\X}f_{0,\bw\vert\bx}(\bw\vert\bx)f_{\bx}(\bx)d\bx}{\int_{\X}f_{\bw\vert\bx}(\bw\vert\bx)f_{\bx}(\bx)d\bx}\right\}f_{0,\bw}(\bw)d\bw}}\\
    &&\hskip -35pt =I_{1}+I_{2}\label{eq:KLbd}.
\ese
\vspace{-5ex}\\
We can further rewrite the first term as 
\vspace{-5ex}\\
\bse
    &I_{1} =\int_{\W_{0}} \log\left\{\frac{\int_{\X}f_{0,\bw\vert\bx}(\bw\vert\bx)f_{0,\bx}(\bx)d\bx}{\int_{\X}f_{0,\bw\vert\bx}(\bw\vert\bx)f_{\bx}(\bx)d\bx}\right\}f_{0,\bw}(\bw)d\bw + \int_{\W_{0}^{c}} \log\left\{\frac{\int_{\X}f_{0,\bw\vert\bx}(\bw\vert\bx)f_{0,\bx}(\bx)d\bx}{\int_{\X}f_{0,\bw\vert\bx}(\bw\vert\bx)f_{\bx}(\bx)d\bx}\right\} f_{0,\bw}(\bw)d\bw.
\ese
\vspace{-5ex}\\
We can then choose a compact set $\W_{0}$ such that  
\vspace{-5ex}\\
\be\label{eq: I1part1}
    &\int_{\W_{0}^{c}} \log\left\{\frac{\int_{\X}f_{0,\bw\vert\bx}(\bw\vert\bx)f_{0,\bx}(\bx)d\bx}{\int_{\X}f_{0,\bw\vert\bx}(\bw\vert\bx)f_{\bx}(\bx)d\bx}\right\} f_{0,\bw}(\bw)d\bw\nonumber\\
    &\quad\leq\int_{\W_{0}^{c}}  \log\left\{\frac{\int_{\X}f_{0,\bw\vert\bx}(\bw\vert\bx)f_{0,\bx}(\bx)d\bx}{\inf_{\bx}f_{0,\bw\vert\bx}(\bw\vert\bx)}\right\} f_{0,\bw}(\bw)d\bw \leq\epsilon/4.
\ee
\vspace{-5ex}\\
Existence of such an $\W_{0}$ follows from Assumption \ref{assmp: KL w}.

Under Assumption~\ref{assmp: func}, for $\U=\{f_{\bx}:\|f_{\bx}- f_{0,\bx}\|<\epsilon/(8M)\}$, we have
\vspace{-5ex}\\
\be
    &\left|\int f_{0,\bw\vert\bx}(\bw\vert\bx)f_{\bx}(\bx)d\bx-\int f_{0,\bw\vert\bx}(\bw\vert\bx)f_{0,\bx}(\bx)d\bx\right|\nonumber\\ %
    &\leq eM\int\left|f_{\bx}(\bx)- f_{0,\bx}(\bx)\right| d\bx= eM\|f_{\bx}- f_{0,\bx}\| \leq e\epsilon/8.
\ee
\vspace{-5ex}\\
When $\epsilon<4$, we have 
\vspace{-5ex}\\
\be
    \left|\frac{\int f_{0,\bw\vert\bx}(\bw\vert\bx)f_{0,\bx}(\bx)d\bx}{\int f_{0,\bw\vert\bx}(\bw\vert\bx)f_{\bx}(\bx)d\bx}-1\right|\leq \frac{\epsilon/8}{1-\epsilon/8}\leq
\epsilon/4.\label{eq: I1part2}
\ee
\vspace{-5ex}\\
Combining \eqref{eq: I1part1} and \eqref{eq: I1part2}, we have $I_{1}<\epsilon/{2}$ for $f_{\bx}\in \mathcal{U}$. 
For the second term $I_{2}$, we first note that 
\vspace{-5ex}\\
\bse
\frac{\int_{\X}f_{0,\bw\vert\bx}(\bw\vert\bx)f_{\bx}(\bx)d\bx}{\int_{\X}f_{\bw\vert\bx}(\bw\vert\bx)f_{\bx}(\bx)d\bx} 
= \frac{\int_{\X}\frac{f_{0,\bw\vert\bx}(\bw\vert\bx)}{f_{\bw\vert\bx}(\bw\vert\bx)} f_{\bw\vert\bx}(\bw\vert\bx) f_{\bx}(\bx)d\bx}{\int_{\X}f_{\bw\vert\bx}(\bw\vert\bx)f_{\bx}(\bx)d\bx}
\leq\sup_{\bx}\frac{f_{0,\bw\vert\bx}(\bw\vert\bx)}{f_{\bw\vert\bx}(\bw\vert\bx)}.
\ese
\vspace{-5ex}\\
Since $\bx$ is in a compact compact support, we have $\eE(\bw)=\eE\{\eE(\bw\vert\bx)\}=\eE(\bx)<\infty$.
Given this bounded expectation and
the monotonicity of the logarithm function, 
an application of the dominated convergence theorem 
gives us $I_{2}\rightarrow 0$ as $\btheta'\rightarrow\btheta'_{0}$.
Choose a neighborhood $N_{1}=\{\btheta':\|\btheta'-\btheta'_{0}\|_{\infty}<\delta_{1}\}$ of $\btheta_{01}$ such that for $\btheta' \in N_{1}$, $I_{2}<\epsilon/2$. Combining the results for $I_{1}$ and $I_{2}$, we have KL$\{f_{0,\bw}(\bw),f_{\bw}(\bw)\}<\epsilon$ when $\btheta'\in N_{1}$ and $f_{\btheta'',\bx}\in\mathcal{U}$, where $\btheta=\btheta'\bigcup\btheta''$.

\newpage
Let us define 
\vspace{-5ex}\\
\bse
& f_{0,K,\bx}(\bx)=\det(\bR_{0,\bx})^{-d/2}\exp\left\{-\by_{\bx}\trans(\bR_{0,\bx}^{-1}-\bI_{d})\by_{\bx}\right\}\prod_{\ell=1}^{d} f_{0,K,x,\ell}(x_{\ell})\\
&\text{with}~~ f_{0,K,\bx,\ell}(x_{\ell})=\sum_{k=1}^{K}\pi_{\ell,k}\TN(x_{\ell}\vert\mu_{k},\sigma_{k},[A,B])~~\text{and}~~{F_{0,K,x,\ell}(x_{\ell})=\Phi(y_{x,\ell})},~~\ell=1,\dots,d.
\ese
\vspace{-5ex}\\
By universal approximation theorem, there exists $K_{\epsilon}$ such that for all $K>K_{\epsilon}$, we can bound the approximation error due to a mixture $f_{0,K,x,\ell}$ by $\epsilon/(16M\xi d)$ for each $\ell$, 
where $\xi=\int_{\X}\frac{\MVN(\by\mid 0,\bR_{0,\bx})}{\MVN(\by\mid 0,\bI_{d})} d\bx$. 
By Cauchy-Squartz inequality, we have for positive valued functions $a(t)$ and $b(t)$ of $t$, $\int a(t)b(t)dt\leq (\int \sqrt{a(t)b(t)}dt)^2\leq\int a(t)dt\int b(t)dt$. Using that we have the following, 
\vspace{-5ex}\\
\bse
\|f_{0,K,\bx}- f_{0,\bx}\|\leq d\max_{\ell}\|f_{0,K,x,\ell}-f_{0,x,\ell}\| \int_{\X}\frac{\MVN(\by\mid 0,\bR_{0,\bx})}{\MVN(\by\mid 0,\bI_{d})}d\bx\leq (\epsilon/16M\xi)\xi\leq \epsilon/(16M).
\ese
\vspace{-5ex}\\
We thus have 
\vspace{-5ex}\\
\bse
\|f_{\bx}- f_{0,\bx}\|\leq \|f_{0,K,\bx}- f_{0,\bx}\| + \|f_{\bx}- f_{0,K,\bx}\| \leq \epsilon/(16M) + \|f_{\bx}- f_{0,K,\bx}\|.
\ese
\vspace{-5ex}\\
Here we consider the same set of locations and scales for all the components.
In the worst case, the mixture approximation of each component is equivalent to an approximation with $K/d$ many truncated normal distributions.
Due to continuity, there exists $\delta_{1}, \delta_{2},\delta_{3}, \delta_4$ such for $\|\mu-\mu_{0}\|_{\infty}<\delta_{1}$, $\|\sigma-\sigma_{0}\|_{\infty}<\delta_{2}$, $\|\pi-\pi_{0}\|_{\infty}<\delta_{3}$, $\|\bR_{\bx}-\bR_{0,\bx}\|_{\infty}<\delta_4$, we have $\|f_{\bx}- f_{0,K,\bx}\|<\epsilon/(16M)$ and take $K>K_{\epsilon}$.
Therefore, 
\vspace{-5ex}\\
\bse
\hspace*{-1cm} \Pi(\U)>\Pi(\|\mu-\mu_{0}\|_{\infty}<\delta_{1})\Pi(\|\sigma-\sigma_{0}\|_{\infty}<\delta_{2})\Pi(\|\pi-\pi_{0}\|_{\infty}<\delta_{3})\Pi(\|\bR_{\bx}-\bR_{0,\bx}\|_{\infty}<\delta_4)\Pi(K>K_{\epsilon}).
\ese
\vspace{-5ex}\\
The approximation error for a H\"older smooth true function with regularity $\iota$ using $K$ many B-spline bases is bounded by $K^{-\iota}$. 
Thus we have $\|\kappa_{0,l}-\kappa_{l}\|_{\infty}\leq K_{\kappa}^{-\iota}+\|\bbeta_{0,\kappa,\ell}-\bbeta_{\kappa,\ell}\|_{\infty}$.
Similarly $\|s^{2}_{0}-s^{2}\|_{\infty}\leq K_{s}^{-\iota'}+\|\bbeta_{0,s}-\bbeta_{s}\|_{\infty}$.
Take $K_{\kappa}>(\delta_{1}/4)^{-\iota}, K_{s}>(\delta_{1}/4)^{-\iota}$. 
Hence 
\vspace{-5ex}\\
\bse 
\hspace*{-0.75cm} \Pi(N_{1})>\{\Pi(K_{\kappa}>(\delta_{1}/4)^{-\iota})\}^d\prod_{l=1}^d\Pi(\|\bbeta_{0,\kappa,l}-\bbeta_{\kappa,l}\|_{\infty}<\delta_{1}/4)\Pi(K_{s}>(\delta_{1}/4)^{-\iota'})\Pi(\|\bbeta_{0,s}-\bbeta_{s}\|_{\infty}<\delta_{1}/4).
\ese
\vspace{-5ex}\\
Hence, the KL-support condition is satisfied noting that $\Pi(N_{1})>0$ and $\Pi(\mathcal{U})>0$.

The KL-support condition ensures weak consistency of the posterior distribution \citeplatex{ghosh2003bayesian}. 
To establish strong consistency for the posterior distribution of $f_{\bw}$, 
we apply Theorem 2 of \citelatex{ghosal1999posterior}, 
stated below for easy reference.

\begin{Thm}[\citelatex{ghosal1999posterior}]
Let $\Pi$ be a prior on $\mathscr{F}$. Suppose $f_{0}\in\mathscr{F}$ is in the KL support of $\Pi$ and let $U=\{f:\|f-f_{0}\|<\epsilon\}$. If there is a $\delta<\epsilon/4, c_{1}, c_{2}>0,\beta<\epsilon^{2}/8$ and $\mathscr{F}_n\subset \mathscr{F}$ such that, for all $n$ large:
\begin{itemize}
    \item[(i)] $\Pi(\mathscr{F}_n^c)<c_{1}\exp(-nc_{2}),$ and
    \item[(ii)] $J(\delta,\mathscr{F}_n)<n\beta$,
\end{itemize}
then $\Pi(U\vert X_{1},\ldots, X_n)\rightarrow 1$ a.s. $P_{f_{0}}$. 
\label{ghosalthm}
\end{Thm}
The constants $\delta, c_{1}, c_{2}, \beta$ and the subset $\mathscr{F}_n$ may depend on $\epsilon$.
In the above theorem, the strong consistency is handled based on conditions in terms of $L_{1}-$metric entropy $J(\delta,\mathscr{F}_n)$.
We are showing strong consistency for $f_{\bw}$ here. Thus our $U=\{f_{\bw}:\|f_{\bw}-f_{0,\bw}\|<\epsilon\}$.

Consider the sieve in the parameter space
\vspace{-5ex}\\
\bse
    \mathscr{H}_n=&\{\btheta:\|\bbeta_{\kappa}\|_{\infty}<M_{1n}, \bbeta_{s}\in[m_{2n},M_{2n}]^{K_s},\|\mu\|_{\infty}<M_{3n},m_{4n}<\|\sigma\|_{\infty}<M_{4n}, \\&K_{\kappa}\leq N_{1n}, K_{s}\leq N_{2n}, K\leq N_{3n}\}~~\text{and}~~ \mathscr{F}_n=\{f_{\btheta,\bw}:\btheta\in\mathscr{H}_n\}.
\ese
\vspace{-5ex}\\
Let us also define 
\vspace{-5ex}\\
\bse
& \mathscr{H}_{1n}=\{\btheta':\|\bbeta_{\kappa}\|_{\infty}<M_{1n}, \|\bbeta_{s}\|_{\infty}<M_{2n}, K_{\kappa}\leq N_{1n}, K_{s}\leq N_{2n}\},\\
&\text{and}~~\mathscr{F}_{1n}=\{f_{\btheta_{1},\bw\vert \bx}:\btheta_{1}\in\mathscr{H}_{1n}\},\\
&\mathscr{H}_{2n}=\{\btheta'':\|\mu\|_{\infty}<M_{3n},m_{4n}<\|\sigma\|_{\infty}<M_{4n}, K\leq N_{3n}\},\\
&\text{and}~~\mathscr{F}_{2n}=\{f_{\btheta_{2},\bx}:\btheta_{2}\in\mathscr{H}_{2n}\}.
\ese
\vspace{-5ex}\\
We have
\vspace{-5ex}\\
\be
    \|f_{1,\bw}-f_{2,\bw}\|&\leq\int\int f_{1,\bw|\bx}|f_{1,\bx}-f_{2,\bx}|d\bw d\bx + \int\int f_{2,\bx}|f_{1,\bw|\bx}-f_{2,\bw|\bx}|d\bw d\bx\nonumber\\
    &\quad\leq\int\int f_{1,\bw|\bx}|f_{1,\bx}-f_{2,\bx}|d\bw d\bx + \int f_{2,\bx}\int |f_{1,\bw|\bx}-f_{2,\bw|\bx}|d\bw d\bx\nonumber\\&\qquad\leq\|f_{1,\bx}-f_{2,\bx}\|+\sup_{\bx}\|f_{1,\bw|\bx}-f_{2,\bw|\bx}\|\nonumber\\
    &\quad\qquad\leq2d_{H}(f_{1,\bx},f_{2,\bx})+\sup_{\bx}\|f_{1,\bw|\bx}-f_{2,\bw|\bx}\|\label{eq:ineq1}.
\ee
\vspace{-5ex}\\
Here $d_{H}(f_{1},f_{2})$ stands for the Hellinger distance between two densities $f_{1}$ and $f_{2}$. 
The $L_{1}$ distance between two normal densities can be bounded as $\|\Normal(\mu_{1},\sigma_{1})-\Normal(\mu_{2},\sigma_{2})\|\leq C_{1}\frac{|\mu_{1}-\mu_{2}|}{\sigma_{1}\wedge\sigma_{2}}$ for some constant $C_{1}$. 
To bound the $L_{1}$ distance between two MvMF densities, we first apply Pinsker's inequality. 
We have from the results of MvMF that $\eE(\bQ_{\ell,\ell})=\frac{\partial\log(M(\bF))}{\partial\bf_{\ell}}$, where $\bQ\sim\MvMF(\bF)$ with $\bF=\diag(\bFf)$. Since $\eE(\bQ_{\ell,\ell})\leq 1$, we have $\frac{\partial\log(M(\bF))}{\partial f_{\ell}}\leq 1$. 
Thus, in light of the mean value theorem, we have $|\log(M(\bF_{1}))-\log(M(\bF_{2}))|\leq \|\bF_{1}-\bF_{2}\|_{\infty}$.
Hence, we have
\vspace{-5ex}\\
\bse
  &\|\MvMF(\bF_{1}), \MvMF(\bF_{2})\|^{2} \leq 2\textrm{KL}(\MvMF(\bF_{1}), \MvMF(\bF_{2}))\\
  &\quad\leq 2|\log(M(\bF_{1}))-\log(M(\bF_{2}))|+2\|\bF_{1}-\bF_{2}\|_{\infty} \leq 4\|\bF_{1}-\bF_{2}\|_{\infty}.
\ese
\vspace{-5ex}\\
As $f_{\bw|\bx}$ is product of two  densities MvMF and log-normal, combining the above bounds, we have
\vspace{-5ex}\\
\be
    \sup_{\bx}\|f_{1,\bw|\bx}-f_{2,\bw|\bx}\|^2&\leq C_{1}^2\frac{\|s_{1}^{2}-s_{2}^{2}\|_{\infty}^2}{m_{2n}^2}+4\|\bF_{1}-\bF_{2}\|_{\infty} \\
    &\leq C_{1}^2\frac{\|\bbeta_{1s}-\bbeta_{2s}\|_{\infty}^2}{m_{2n}^2}+4\|\bbeta_{1\kappa}-\bbeta_{2\kappa}\|_{\infty}.
\ee
\vspace{-5ex}\\
Using Theorem 1 of \citelatex{genovese2000rates}, we have following bound for the bracketing number with some constant $c$ as $\exp(J_{1}(\delta_{1},\mathscr{F}_{2n}, d_H))\leq \sum_{k=1}^{N_{3n}}cM_{3n}^k\left(\frac{M_{4n}}{m_{4n}}\right)^{2k}\frac{1}{\delta_{1}^{3k-1}}$, where $J_{1}$ stands for logarithm of minimum number of brackets of size $\delta_{1}$ required to cover $\mathscr{F}_{2n}$.
For sufficiently large $n$, we have $M_{3n}>1$ and $\frac{M_{4n}}{m_{4n}}>1$. 
Thus for $\delta_{1}<1$, we have $\exp(J_{1}(\delta_{1},\mathscr{F}_{2n}, d_{H}))\leq N_{3n}M_{3n}^{N_{3n}}\left(\frac{M_{4n}}{m_{4n}}\right)^{2N_{3n}}\frac{1}{\delta_{1}^{3N_{3n}-1}}$. 
It is easy to see that $J(\delta/4,\mathscr{F}_{2n}, d_{H})\leq J_{1}(\delta/2,\mathscr{F}_{2n}, d_{H})$ as balls of radius of $\delta/4$ can be covered by brackets of length $\delta/2$.
To calculate the covering number, we first note that from \eqref{eq:ineq1} we have
\vspace{-5ex}\\
\be
    & \hspace{-1.5cm} J(\delta,\mathscr{F}_n, d)\leq J(\delta/2,\mathscr{F}_{1n}, d)+J(\delta/4,\mathscr{F}_{2n}, d_{H})\leq J(\delta/2,\mathscr{F}_{1n}, d)+J_{1}(\delta/2,\mathscr{F}_{2n}, d_{H}),\\
    & J(\delta/2,\mathscr{F}_{1n}, d)\leq J(m_{2n}\delta/C_{1},\{\bbeta_{s}, K_s:\bbeta_{s}\in[m_{2n},M_{2n}]^{K_s},K_{s}\leq N_{2n}\}, \|\cdot\|_{\infty})\nonumber\\
    &\quad+J(\delta/16,\{\bbeta_{\kappa}, K_{\kappa}:\|\bbeta_{\kappa}\|_{\infty}<M_{1n},K_{\kappa}\leq N_{1n}\}, \|\cdot\|_{\infty})\nonumber\\
    &\quad\leq N_{2n}\log\{3C_{1}N_{2n}M_{2n}/(m_{2n}\delta)\}+N_{1n}\log(48N_{1n}M_{1n}/\delta),\\
    & \hspace{-1.5cm}  J_{1}(\delta/2,\mathscr{F}_{2n}, d_{H})\leq \log(N_{3n})+N_{3n}\log(M_{3n})+2N_{3n}\log(M_{4n}/m_{4n})+(3N_{3n}-1)\log(2/\delta).
\ee    
\vspace{-5ex}\\
The prior probability of the sieve-complement can be bounded as,
$\Pi(\mathscr{H}_n^c)<\Pi(\bbeta_{\kappa}\notin[0,M_{1n}]^{N_{1n}})+\Pi(\bbeta_{s}\notin[m_{2n},M_{2n}]^{N_{2n}})+\Pi(\mu\notin[-M_{3n},M_{3n}]^{N})+\Pi(\sigma\notin[m_{4n},M_{4n}]^{N})+\Pi(K_{\kappa}>N_{1n})+\Pi(K_{s}>N_{2n})+\Pi(K>N_{3n})$.
Combining all of these prior probabilities, we get
\vspace{-5ex}\\
\bse
    & \Pi(\mathscr{H}_n^c)<N_{1n}\exp(-R_{1}M_{1n}^{t1})+N_{2n}\exp(-R_{2}M_{2n}^{t2})+N_{3n}\{\exp(-R_{3}M_{3n}^{t3})+\exp(-R_4M_{4n}^{t3})\}\\
    &~~~+\exp\{-N_{1n}(\log N_{1n})^{b_{\kappa 3}}\}+\exp\{-N_{2n}(\log N_{2n})^{b_{s 3}}\}+\exp\{-N_{3n}(\log N_{3n})^{b_{3}}\},
\ese
\vspace{-5ex}\\
for some constants $R_{1}, t_{1}, R_{2}, t_{2}, R_{3}, t_{3}$.
As per the requirements in Theorem~\ref{ghosalthm}, for $\delta<\epsilon/4$ and $\beta<\epsilon^{2}/8$ we need 
\vspace{-5ex}\\
\bse
    &N_{2n}\log\{3C_{1}N_{2n}M_{2n}/(m_{2n}\delta)\}+N_{1n}\log(48N_{1n}M_{1n}/\delta)+\log(N_{3n})+N_{3n}\log(M_{3n})\\&+2N_{3n}\log(M_{4n}/m_{4n})+(3N_{3n}-1)\log(2/\delta) < n\beta,
~~~\text{and}\\
    &N_{1n}\exp(-R_{1}M_{1n}^{t1})+N_{2n}\exp(-R_{2}M_{2n}^{t2})+N_{3n}\{\exp(-R_{3}M_{3n}^{t3})+\exp(-R_4M_{4n}^{t3})\}\\
    &+\exp\{-N_{1n}(\log N_{1n})^{b_{\kappa 3}}\}+\exp\{-N_{2n}(\log N_{2n})^{b_{s 3}}\}+\exp\{-N_{3n}(\log N_{3n})^{b_{3}}\} < c_{1}\exp(-nc_{2}).
\ese
\vspace{-5ex}\\
If we choose $N_{1n}, M_{1n}, N_{2n}, M_{2n}, N_{3n}, M_{3n},$ and $M_{4n}$ as positive polynomial of $n$ depending on the constants in above equations and choose $m_{2n}$ and $ m_{4n}$ as negative polynomials of $n$, we are done.

\newpage
\section{Additional Figures} \label{sec: add figs}

\begin{figure}[!htbp]
    \centering
    \includegraphics[width = 1\textwidth]{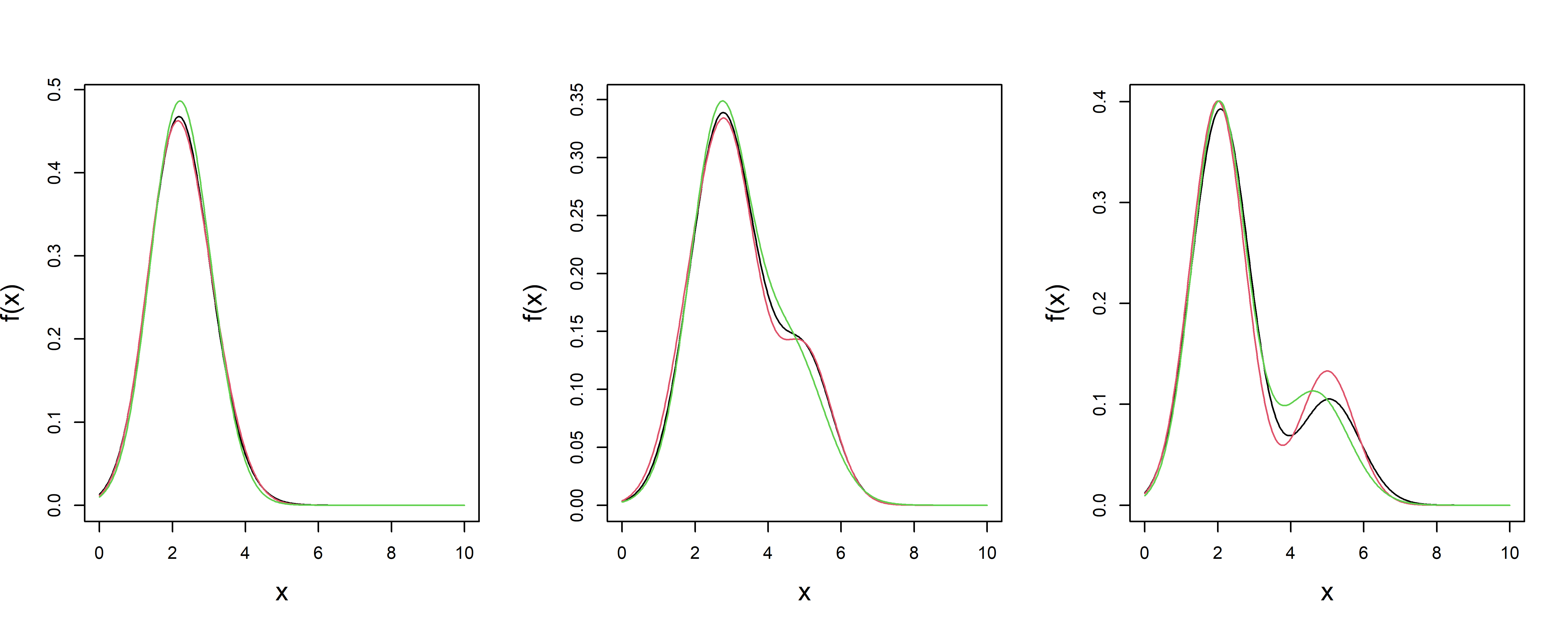}
    \caption{Estimated densities for simulated data with $n=1000$ subjects and $m_{i}=3$ replicates per subject when the true data generating process follows the proposed structure from model (4) in the main paper. 
    We present the estimates from our proposed method (in black) and the method of \cite{sarkar2021bayesian} (in green) along with the truth (in red).}
    \label{fig:simplot11}
\end{figure}

\begin{figure}[!htbp]
    \centering
    \includegraphics[width = 1\textwidth]{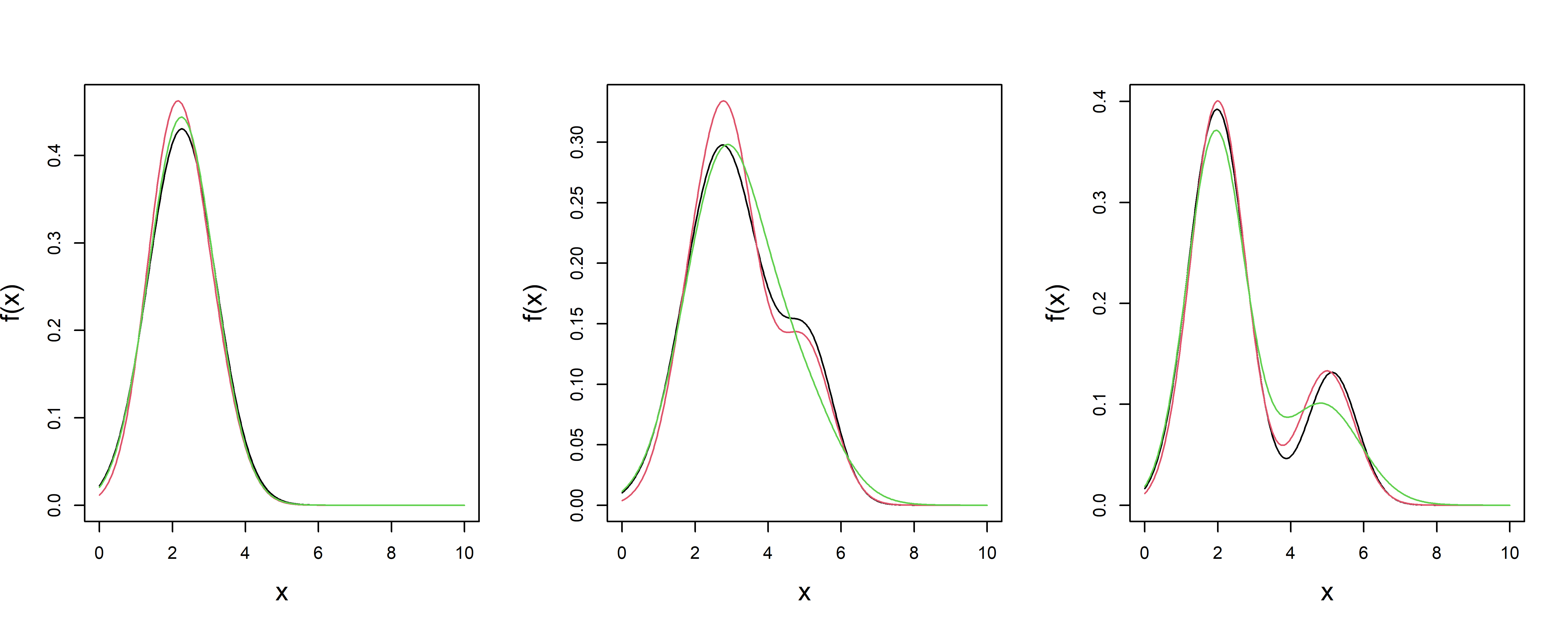}
    \caption{Estimated densities for simulated data with $n=1000$ subjects and $m_{i}=3$ replicates per subject when the true data generating process follows the additive model from \cite{sarkar2021bayesian}. 
    We present the estimates from our proposed method (in black) and the method of \cite{sarkar2021bayesian} (in green) along with the truth (in red).}
    \label{fig:simplot12}
\end{figure}

\bibliographystylelatex{natbib}
\bibliographylatex{main}

\end{document}